\documentclass[12pt]{article}
\pdfoutput=1

\usepackage{amsmath,amssymb,amscd}

\usepackage{listings}
\usepackage{caption}
\usepackage{dsfont}
\usepackage{slashed}
\usepackage{color}
\usepackage{ulem}
\usepackage[pdftex]{graphicx}
\usepackage{epstopdf}
\usepackage{subfigure}
\usepackage{epsfig}
\usepackage{listings}
\usepackage{caption}
\usepackage{cite}
\usepackage{multirow}
\usepackage{comment}
\usepackage{hyperref}

\newcommand{\e}{\text{e}}
\newcommand{\n}{\text{n}}
\newcommand{\D}{\text{D}}

\newcommand{\beq}{\begin{equation}}
\newcommand{\eeq}{\end{equation}}
\newcommand{\nbea}{\begin{align*}}
\newcommand{\neea}{\end{align*}}
\newcommand{\nbeq}{\begin{equation*}}
\newcommand{\neeq}{\end{equation*}}

\newcommand{\hinf}{H_{\rm inf}}
\newcommand{\hinfinv}{H_{\rm inf}^{-1}}

\usepackage{multirow}
\usepackage{array}
\newcolumntype{M}[1]{>{\centering\arraybackslash}m{#1}}
\newcolumntype{N}{@{}m{0pt}@{}}

\begin{document}

\baselineskip=21pt

\begin{center}

{\large {\bf Primordial Black Holes from Spatially Varying Cosmological Constant
Induced by Field Fluctuations in Extra Dimensions}}

\vskip 0.2in
\vskip 0.2in

{\bf Arkady A. Popov}\textsuperscript{a}, 
{\bf Sergey G.~Rubin}\textsuperscript{a,b} and~
{\bf Alexander S.~Sakharov}\textsuperscript{c,d}~

\vskip 0.2in
\vskip 0.2in

{\small {\it

\textsuperscript{a}{\mbox N.~I.~Lobachevsky Institute of Mathematics and Mechanics, Kazan  Federal  University\\
420008, \mbox{Kremlevskaya  street  18,  Kazan,  Russia}}\\
\vspace{0.25cm}
\textsuperscript{b}{\mbox National Research Nuclear University MEPhI (Moscow Engineering Physics Institute)}\\
\mbox{Kashirskoe Shosse 31, 115409 Moscow, Russia}\\
\vspace{0.25cm}
\textsuperscript{c}{\mbox Physics and Astronomy Department, Manhattan College}\\
\mbox{4513 Manhattan College Parkway, Riverdale, NY 10471, United States of America}\\
\vspace{0.25cm}
\textsuperscript{d}{\mbox Experimental Physics Department, CERN, CH-1211 Gen\`eve 23, Switzerland}\\
\vspace{0.25cm}
}
}

\vskip 0.2in

{\bf Abstract}

\end{center}

\baselineskip=18pt
\noindent


The origin and evolution of supermassive black holes (SMBHs) in our universe
have sparked controversy. In this study, we explore the hypothesis that some of
these black holes may have seeded from the direct collapse of dark energy domains with density significantly higher than the surrounding regions. The mechanism of the origin of such domains relies on the inflationary evolution of a scalar field acting in $D$ dimensions, which is associated with the cosmological constant in our four-dimensional spacetime manifold. Inner space quantum fluctuations of the field during inflation are responsible for the spatial variations of the dark energy density in our space. This finding holds particular significance, especially considering recent evidence from pulsar timing array observations, which supports the existence of a stochastic gravitational wave background consisting of SMBH mergers.

\vskip 3mm

\noindent Keywords: Primordial black hole; Supermassive black hole; Cosmological constant; \\ Dark energy; Extra dimensions; Inflation
\vskip 5mm

\today

\vskip 10mm

\section{Introduction}
\label{intro}

Primordial black holes (PBHs) have been extensively studied over the
decades~\cite{1967SvA....10..602Z, 1971MNRAS.152...75H, 1974MNRAS.168..399C,
1980PhLB...97..383K, 1985UsFiN.145..369P, 2010RAA....10..495K,
2019EPJC...79..246B, 2018CQGra..35f3001S,
2021PhRvD.103h3518G,
Caretal20}
and offer a scenario with the potential to leave distinct imprints on cosmic history.
Depending on the ratio of their abundance relative to the overall dark matter (DM),
$f_{\rm PBH} = \Omega_{\rm PBH}/\Omega_{\rm DM}$, the range of possible PBH masses $M_{\rm PBH}$ spans a wide spectrum, including PBHs of small masses \cite{2021PhRvD.103h3518G} which have undergone scrutiny through various observations (for comprehensive reviews, refer to~\cite{2018CQGra..35f3001S, Caretal20}).
Additionally, since PBHs formed during the early stages of the Universe, they have the capacity to develop
bound binaries via multiple mechanisms~\cite{2018CQGra..35f3001S,pbhMech2,pbhMech3}. As these binaries become close,
they emit gravitational waves (GWs) continuously until a final dramatic burst occurs at the point of their ultimate merger.
Notably, for black holes of stellar mass, such mergers have already been detected by ground-based interferometers~\cite{LVK1,LVK2}.
Moreover, it is plausible that several of these observed mergers might be attributed to the coalescence of PBHs~\cite{LVKdetails1,LVKdetails2,LVKdetails3,LVKdetails4,LVKdetails5,LVKdetails6}.

PBHs with masses exceeding $10^2M_{\odot}$ hold particular significance due to their impact on the growth of massive objects during the
evolution of the early Universe. Notably, it is well-established that galactic nuclei host supermassive black holes (SMBHs) with masses
surpassing $10^6\, M_{\odot}$~\cite{smbh1,smbh2,smbh3}. It has been theorized that PBHs could be their progenitors, achieving such masses
through processes like merging, accretion~\cite{acr1,acr2,acr3,acr4,acr5,acr6}, or the direct collapse of primordial
fluctuations~\cite{fluc1,fluc2}. In the latter scenario, SPBHs are constrained to constitute less than O(0.1\%) of dark matter (DM). As they
have been present since the dawn of the matter-dominated era, they can serve as cosmic seeds, enhancing galaxy formation~\cite{galform1,galform2}.
Furthermore, different observations have provided evidence for the existence of
intermediate-mass black holes (IMBHs) with masses ranging from $10^4\, M_{\odot}$ to $10^6\, M_{\odot}$~\cite{IMBH_1}.
Additionally, a subdominant fraction of dark matter may consist of immensely massive PBHs, exceeding $10^{12}\, M_{\odot}$~\cite{smbhdm1},
capable of traversing the intergalactic medium.

SMBHs may be responsible for the generation of early galaxies reported by JWST~\cite{jwst1}.
They can bind in binary systems which leads to late time merging and radiation of gravitational waves in the nHz frequency range that are detectable
by pulsar timing array (PTA) experiments~\cite{pta1,pta2,pta3,pta4,pta5,pta6,pta7}. 
The results from the PTA observations have been extensively
analyzed and interpreted in numerous studies, including recent ones such as~~\cite{Valli:2024nbj,
Alonso-Alvarez:2024gdz,Buchmuller:2024zzk,Winkler:2024olr,Conaci:2024tlc,
Choudhury:2024one,Padmanabhan:2024nvv,Hu:2023oiu,Lacy:2023kbb,Eichhorn:2023iab,Zhang:2023jrk,Sato-Polito:2023gym,Liu:2023pvq,Ellis:2023iyb,
Huang:2023klk,Bromley:2023yfi,Davis:2023vyy,Harris:2023xab,Koo:2023gfm,Ramazanov:2023eau,DOrazio:2023rvl,Kasai:2023qic,Stamou:2023vft,
Dolgov:2023ijt,Huang:2023nes,Evans:2023jia,Gardiner:2023zzr,Serra:2023kkk,Cyr:2023pgw,InternationalPulsarTimingArray:2023mzf,
Flores:2023dgp,Kawasaki:2023rfx,Ellis:2023oxs,Bhaumik:2023wmw,Babichev:2023pbf,Buchmuller:2023aus,Gouttenoire:2023bqy,Wu:2023hsa,Zhang:2023lzt,
Gouttenoire:2023nzr,Bi:2023tib,Broadhurst:2023tus,Huang:2023chx,Ellis:2023tsl,
Ellis:2023dgf,Addazi:2023jvg,Furusawa:2023fwl,Chen:2024fir,Li:2024psa,
Liu:2023tmv,Chen:2023bms,Kitajima:2023kzu,LISACosmologyWorkingGroup:2023njw,Liu:2023hpw,Chung:2023xcv,King:2023ayw,Zhu:2023lbf,Liu:2023pau,
Ahmadvand:2023lpp,Zhang:2023nrs,Cannizzaro:2023mgc,Jin:2023wri,Servant:2023mwt,Li:2023bxy,Gouttenoire:2023ftk,Blasi:2023sej,
Vagnozzi:2023lwo,Franciolini:2023pbf,Athron:2023mer,Zeng:2023jut} and earlier analyses such as~\cite{Gonzalez:2022mcx,Blasi:2022ayo,Chattopadhyay:2022fwa,Wu:2022tpe,
Ferreira:2022zzo,Wu:2022stu,Ashoorioon:2022raz,Sun:2021yra,Babichev:2021uvl,Benetti:2021uea,Kirillov:2021qjz,
Brandenburg:2021tmp,Sakharov:2021dim}, which are related to the previously published NANOGrav signal evidence~\cite{NANOGrav:2020bcs}.
These interpretations and effects may independently explain the PTA gravitational wave signal.
They can also be considered in combination with the modeling of gravitational waves originating
from supermassive black hole binaries (SMBHBs).

In this paper, we propose and validate a mechanism for the formation of PBHs based on the generation of specially varying cosmological constants, which may be generic for theories with compact extra dimensions. Considering extra dimensions allows us to examine fluctuations of fields within the internal space during inflation in addition to fluctuations of ordinary scalar fields. However, the fate of these field fluctuations differs significantly from those associated with conventional four-dimensional scalar fields.
While conventional field fluctuations transform their energy density into radiation during the FRW stage through decay into other particle-like species, the energy density associated with scalar fields within the internal space remains unchanged, effectively stored within the scalar field itself, manifesting as a cosmological constant. Fluctuations of the scalar field within the internal space manifest as spatial variations of the local $\Lambda$ term. Domains containing an extraordinarily high cosmological constant may collapse into PBHs.
After the end of inflation, the horizon expands and the particle energy density decreases, approaching its present-day value. Simultaneously, the energy density associated with the $\Lambda$ term, being dependent on the Hubble parameter, also decreases over time, converging to its present-day value, which equals the dark energy density. There must exist a moment in time when both the energy density of matter and the energy density associated with the $\Lambda$ term are equal.
Evidently, since the value of $\Lambda$ varies across space coordinates, this equality primarily arises within the densest regions, implying the existence of a density contrast close to unity in domains with high values of $\Lambda$. Once such a domain becomes encompassed by the cosmological horizon, it may evolve into a PBH. We assert that PBHs formed through the proposed mechanism are cosmologically feasible candidates for seeding SMBHs and explaining the observed IMBHs.

The flexible metrics characterizing extra dimensions constitute a continuous set of static classical solutions derived from the generalized Einstein igat \cite{2007PhLB..651..224N, Nojiri_2017}, and they share fixed Lagrangian parameters. 
This approach, distinct from the brane world model, renders the extra dimensions invisible due to their small size. However, unlike Kaluza--Klein geometries, these dimensions exhibit inhomogeneity. The concept of such geometries was initially introduced in \cite{Rubin:2015pqa} with further discussion in \cite{2017JCAP...10..001B}. Subsequent research, as presented in \cite{Rubin:2016ude}, applied a top-down approach to elucidate observed physical laws. It demonstrated that incorporating quantum corrections to initial parameters established at high energies eases the renormalization procedure.

Investigating the evolution of extra field distributions leading to a static state is a crucial endeavor. This aspect has been explored in previous works such as \cite{Bronnikov:2023lej}. The outcomes reveal that the resultant metric and field distribution are contingent upon both model parameters and initial conditions. Notably, the extra-dimensional stationary field distributions evolve in tandem with the energy density across distinct volumes below the horizon, which are replicated during inflation. This particular aspect forms the focal point of our investigation.

Furthermore, our current investigation is grounded in nonlinear $f(R)$ gravity, as extensively discussed in reviews such as \cite{Nojiri:2006ri, Sotiriou:2008rp}. This framework holds significant potential for diverse cosmological implications, with one notably remarkable consequence being the emergence of dark matter \cite{Arbuzova:2021etq}. Several viable $f(R)$ models in 4D space aligning with observational constraints have been proposed in works such as \cite{DeFelice:2010aj, 2014JCAP...01..008B, Sokolowski:2007rd, 2007PhLB..651..224N, Nojiri_2017}.

This paper is structured as follows: In Section~\ref{extradim},
we provide a concise overview of the mathematical setup
employed in the extra-dimensional framework under
consideration. Section~\ref{infl} is dedicated to exploring the
distinct behaviors of fluctuations in our space compared to
those in extra dimensions. Section~\ref{PBH} is focused on
deriving the conditions essential for the formation of PBHs
and estimating their mass spectrum. The conclusions
of our study are summarized in Section~\ref{conclude}.

\section{Static Field Distribution in Internal Space}
 \label{extradim}

The primary objective of this section is to revisit the foundational concepts of extra-dimensional frameworks that give rise to a continuum set of static metric distributions. This issue has been elaborated in our previous papers \cite{Bronnikov:2023lej}, and we refer the reader to them for details.

Consider $f(R)$ gravity with a minimally coupled scalar field $\zeta$ in a $D = 4 + n$-dimensional manifold $M_D = M_4 \times M_n$:
\begin{eqnarray}\label{SD}
	S = \frac{m_{D}^{D-2}}{2}\int_{M_D}  d^{D} X \sqrt{|g_{D}|} \,  \Bigl( f(R)
	+ \partial^{M}\zeta \, \partial_{M}\zeta -2 V(\zeta) \Bigr) ,
\end{eqnarray}
 where $g_{D} \equiv \mbox{Det} g_{MN}$; $M,N =\overline{1,D}; \quad X^A=(x^{\mu},y^a)$; {the coordinate set $x^\mu, \mu=1,2,3,4$ describes the four-dimensional space $M_4$, and the set  $y^a, a=5,6,...,n$} describes the $n$-dimensional manifold $M_n$, which is assumed to be a closed manifold without boundary; $f(R)$ is a function of the
 D-dimensional Ricci scalar $R$; and $m_D$ is the $D$-dimensional Planck mass. Below, we  will work in the units $m_D=1$. Note that the main results of this work hold even for the simplest form of the potential $V$.
\begin{eqnarray}\label{V}
	V(\zeta)=\frac 12 m^2 \zeta^2 .
\end{eqnarray}

   The metric is postulated to have the form
\begin{equation}\label{interval}
ds^2=\e^{2 \gamma(u)}\left[dt^2 -\e^{2Ht}(dv^2+ v^2d\Omega_2^2)\right]
- du^2 -r(u)^2 d \Omega_{n-1}^2.
\end{equation}

    Such a metric ansatz has been extensively studied within the realm of linear \\\mbox{gravity \cite{2000PhRvD..62d4014O,2003PhRvD..68b5013C,2005PhRvD..71h4002S,2005PhRvD..71j4018R}}, particularly in addressing the hierarchy problem \cite{Bronnikov:2023lej,2000PhRvL..84.2564G,2000PhRvL..85..240G,Bronnikov:2007kw}. 
    Our approach is based on the concept of compact extra dimensions. A preliminary investigation suggests that their scale could be as small as $10^{-28}$ cm or even smaller. This implies that the extra dimensions remain invisible to our instruments, and our rulers and clocks do not measure intervals of space and time at a specific value of $u$. Instead, all metric functions, such as the function $e^{\gamma(u)}$, should be averaged over the extra space. The way to achieve this is discussed in \cite{Bronnikov:2023lej} and briefly presented below.

   The equations of motion, see \cite{2007PhLB..651..224N, Nojiri_2017}, represented by
 \begin{eqnarray}\label{EE}
&&-\frac 12 \delta_M^N f (R) + \Big( R^N_M + \nabla_M \nabla^N - \delta^N_M \square \Big) f_R =-T_M^N ,\\
&& f_R = df/dR, \quad T_M^N =(\partial_M \zeta) (\partial^N \zeta) -\frac{1}{2} (\partial_C \zeta) (\partial^C \zeta)  \delta^N_M +V(\zeta) \delta^N_M , 
 \nonumber
\end{eqnarray}
possess a continuum set of solutions just as the differential equations do. We choose those solutions that exhibit homogeneity in the spatial coordinates $x$ and inhomogeneity in the internal coordinates $y$. {We consider only those solutions that refer to the compact extra space. This means that the metric function $r(u)$ must have two zeros. This condition is fulfilled at the coordinates $u_{min}$ and $u_{max}$, i.e., $r(u_{min})=r(u_{max})=0$, which is the result of numerical calculations. These coordinate values depend on additional conditions which are different in different space domains.
}

 The parameterization of these solutions is determined by additional conditions, such as $r'(y=0)$, $r(y=0)$, $\gamma(y=0)$, $\gamma'(y=0)$, $\zeta(y=0)$, and $\zeta'(y=0)$, which are essential for solving the second-order differential equations.

 {
    After integration over the extra-dimensional coordinates, the action \eqref{SD} reduces to the effective action~\cite{Rubin:2015pqa} }
\begin{equation}\label{S_eff}
	S = \dfrac{m_P^{2}}{2} \int\limits_{m_P} d^4 x \sqrt{|g_4|}
        \Bigl(a_{eff} R_4^2 + R_4 + c_{eff} \Bigr). 
 \end{equation}
  The
 term $c_{eff}$ represents the cosmological constant $\Lambda$:
\begin{equation}\label{Lambdac}
\Lambda = -\frac{1}{2} c_{eff},
\end{equation}
assuming the scalar function $\zeta$ is homogeneous in a  3-dimensional space under horizon. This value varies in different space regions due to the fluctuations at the inflationary stage. We are interested in those space domains where the effective parameter $\Lambda$ is considerably large as compared to values in the surrounding space.
  Here, $g_4$ is the determinant of the 4D~metric:
\begin{equation}\label{g4}
        ds_4^2 = g_{4,\mu\nu} dx^\mu dx^\nu = dt^2 - \e^{2Ht}\delta_{ij}dx^i dx^j \, .
\end{equation}
The effective parameters are expressed as follows
\begin{eqnarray}       \label{mPl}
       && m^2_P  =  \mathcal{V}_{n-1} \int_{u_{\min}}^{u_{max}}
                    f_R (R_n) \,\e^{2\gamma}\,r^{n-1}\, du, \nonumber \\
               \label{a_eff}
  &&      a_{eff}  = \frac{\mathcal{V}_{\n-1}}{2m_P^{2}}
                \int_{u_{\min}}^{u_{max}} f_{RR}(R_n) \,\e^{4\gamma}\,r^{n-1}\,du,\\
\label{c_eff}
&&        c_{eff}[\zeta]  =  \frac{\mathcal{V}_{n-1}}{m_P^2} \int_{u_{\min}}^{u_{max}}
\Bigl(f(R_n) - \zeta(u)'^2  -
m^2\zeta(u)^2\Bigr)\,\e^{4\gamma}\,r^{n-1}\, du\, ,  \nonumber
\end{eqnarray}
where $\mathcal{V}_{n-1} = \int d^{n-1} x \sqrt{|{g}_{n-1}|} = \dfrac{2\pi^{n/2}}{\Gamma(n/2)}$ is the volume of $n-1$-dim sphere.

\begin{figure*}[t!]
 \centering
 \includegraphics[width=0.33\textwidth]{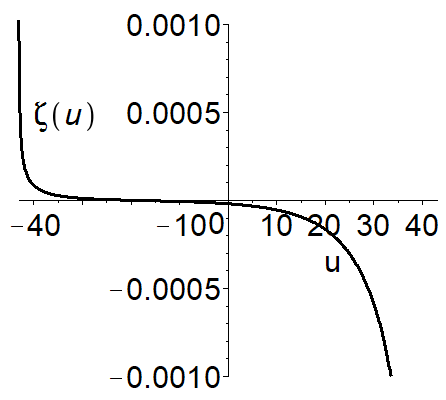}\hspace{0cm}\includegraphics[width=0.33\textwidth]{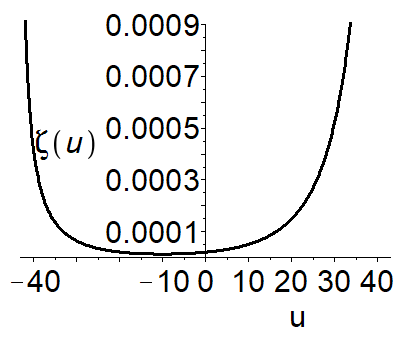}\hspace{0cm}\includegraphics[width=0.33\textwidth]{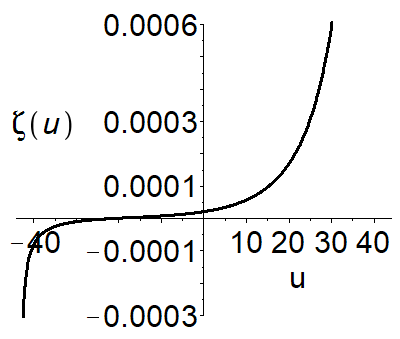}
 \caption{
 Solution of  \eqref{EE} for the following parameters $n=3$, $f(R) =300 R^2 +R +0.002$,
 $H = 0$, $V(\zeta)=0.01\,\zeta^2/2$ and boundary conditions  $r(0) =50$, $\gamma(0)=0$,
$r'(0) = \gamma'(0) = R'(0) = 0$, \ $u_{max} = u_{min} \simeq 43.178$, (\textbf{a}) $R(0) \simeq 0.00396$,  $\zeta(0)=-2\times 10^{-5}$, $\zeta'(0) = 6\times 10^{-8}$, (\textbf{b}) $R(0) \simeq 0.00395$,  $\zeta(0)=2\times 10^{-5}$, $\zeta'(0) = 1.5\times 10^{-6}$, (\textbf{c})  $R(0) \simeq 0.00396$,  $\zeta(0)=-2\times 10^{-6}$, $\zeta'(0) = 2\times 10^{-6}$. The parameter $u$ is expressed in $\D$-dimensional Planck units.
\label{f1}
 }
 \end{figure*}

 The right-hand side of Equation \eqref{mPl} is expressed in units where $m_D=1$. This relation allows us to articulate the D-dimensional Planck mass in terms of the four-dimensional Planck mass $m_P$. In this context, we assume that the functions $\gamma(u),\ r(u),\ \zeta(u),\ R(u)$ constitute a specific solution to the system \eqref{EE}, with details available in \cite{Rubin:2015pqa} for a specific value of the Hubble parameter $H$. Figure \ref{f1} illustrates some examples of static distributions.
This approximation remains valid during the inflationary period and at the present time, particularly when the Hubble parameter remains nearly constant.

Our comprehension of the specific value of the energy density, denoted as $\rho_{\Lambda}=\Lambda m_P^2/(8\pi)= - c_{eff} m_P^2/(16\pi)$, is quite limited. Observational constraints provide an upper limit of approximately $10^{-123}m_P^2$ at the present time. Understanding this
density during inflation is even more uncertain, with the sanity bound being $\rho_{\Lambda}(H\simeq 10^{-6}m_P)\ll H^2\simeq 10^{-12}m_P^2,$ implying its negligible impact on the inflation rate. Post-inflation, considering the variation in the Hubble parameter becomes crucial, and establishing a connection between $\rho_{\Lambda}$ and this parameter remains elusive.
Multiple factors contribute to the complexity of this issue, including quantum corrections, the influence of other fields, and the effects of averaging after the horizon crossing. 
Furthermore, obtaining an accurate solution to the dynamic equations during the reheating stage appears challenging. The subsequent section is dedicated to a detailed discussion of these aspects.

 Clearly, both the extra dimensions and the scalar field experience fluctuations in the D-dimensional space. Quantum fluctuations during inflation are expected to induce significant deviations from their initial values (distributions), especially during the later stages of inflation when scales much smaller than those relevant for CMB observations exit the inflating Hubble patches. Specifically, fluctuations in the parameter $c_{eff}$ during inflation can lead to spatial variations in the cosmological constant. The cosmological effects of these variations are the primary focus of investigation in the subsequent sections of the paper.
These fluctuations may be substantial, giving rise to domains where the
density of dark energy is significantly higher compared to the surrounding regions.
Such domains could persist for an extended duration, provided gravity exerts a strong influence.

\section{Inflationary Field Dynamics in Extra Space}
\label{infl}

The evolution of the Universe is significantly influenced by field fluctuations during inflation. Following the completion of inflation, the rapid decrease of the Hubble parameter induces vigorous damped fluctuations of the field, ultimately converging asymptotically to one of the minima of its potential. These inhomogeneities, influenced by gravitational effects, give rise to a large-scale structure after the conclusion of the radiation-dominated~stage.

We consider the presence of extra spatial dimensions, allowing for fluctuations in fields within this inner space. The destiny of these field fluctuations differs fundamentally from those inherent to the usual, four-dimensional scalar fields described above. This distinction arises from the fact that stationary distributions of fields constitute a set of measured continuum, as established in the early study~\cite{Rubin:2015pqa}. 

Similar to the situation with usual fields, during the FRW epoch, the asymptotic distribution of the fields in the inner space undergoes variations in causally disconnected regions due to random fluctuations during inflation. However, a significant distinction arises: while the energy density of usual fields is transformed into radiation at the FRW stage through the decay of the fields into other particle-like species, the energy density associated with the scalar fields in the inner space remains in its initial form, being stored in the scalar field.
Therefore, the evolution of the energy density of the fields in the inner space is still governed by the Hubble parameter, resulting in a slower decrease compared to the energy density stored in particle-like species generated from the decay of typical scalar fields, which could exist during the inflationary epoch.
To distinguish between these two kinds of energy densities, we use the notation $\rho$ for the energy stored in the ordinary fields fluctuating in the observable three-dimensional space, which is eventually converted into particle-like species, and $\rho_{\Lambda}$ for the energy remaining stored in the scalar field exhibiting inhomogeneities in the inner space. By choosing the model parameters such that $\rho \gg \rho_{\Lambda}$, we ensure that the impact of the field distribution in the inner space on the expansion rate, as well as on the rate of horizon growth, can be safely neglected.

    After the end of inflation, the horizon expands and the particle energy density $\rho$ decreases, approaching its present-day value $\rho(t_0)$. Simultaneously, $\rho_{\Lambda}$, being dependent on the Hubble parameter, also decreases over time, converging to the present-day value $\rho_{\Lambda} (t_0)$, which equals the dark energy density. Since $\rho_{\Lambda}(t_0)>\rho(t_0)$, there must exist a moment in time, denoted by $t_*$, when both densities are equal, $\rho(t_*)=\rho_{\Lambda}(t_*)$.
Evidently, since the value $\rho_{\Lambda}(t_*)$ varies across space coordinates, this equality primarily arises within the densest regions, implying a density contrast $\delta\rho/\rho \simeq 1$.

\section{Formation of PBHs Induced by Inhomogeneous Cosmological Constants}
\label{PBH}

As indicated in the preceding section, our setup involves the total energy density, which comprises the 4D energy density $\rho$, represented either by a scalar field or particle-like species created after the conversion of this field into radiation at the reheating stage, and $\rho_{\Lambda}$, associated with the energy density emerging from the scalar field distribution in the inner space.
During the inflationary stage and for some period afterward, the Universe was dominated by the 4D energy density,
such that $\rho \gg \rho_{\Lambda}$, while at the present time, $\rho \lesssim \rho_{\Lambda_{\rm obs}}$,
where the current observable value of the dark energy density is given by $\rho_{\Lambda_{\rm obs}} \sim 10^{-123}m_{\rm Pl}^4$.
Therefore, at some moment $t_*$ during the evolution of the Universe,
both types of energy density become equal, resulting in
\beq
\label{condEq1}
\rho(t_*) = \rho_{\Lambda}(t_*)\ .
\eeq
If the equality condition (\ref{condEq1}) occurs within a causally connected
domain, it implies that the density contrast,
expressed in this particular case as
\begin{equation}
\label{densContr1}
\frac{\delta\rho}{\rho}(t_*) \approx \frac{\rho_{\Lambda}(t_*)}{\rho(t_*)+\rho_{\Lambda_{\rm obs}}(t_*)},
\end{equation}
exceeds unity.

Let us consider a scenario in which fluctuations of scalar fields in the internal space during inflation
lead to the formation of a domain with size $R(t_{\rm end})$, determined at the end of inflationary epoch
$t_{\rm end}$, filled with a cosmological constant $\Lambda_1$ that exceeds its observable average value $\Lambda_{\rm obs}$.
After the inflationary period, during the FRW epoch,
the domain of size $R(t_{\rm end})$ undergoes simple conformal stretching due to the expansion of the Universe
\beq
\label{exp1}
R(t)=\frac{a(t)}{a(t_{\rm end})}R(t_{\rm end}),
\eeq
where $a(t)$ is the scale factor. It is evident that at a time $t_1 \gtrsim t_*$,
ensuring $\frac{\delta\rho}{\rho (t_*)} >1$, the domain reaches a radius of $R(t_1)$ as
described by Equation (\ref{exp1}), acquiring the mass $M_1$.
Subsequently, it becomes encompassed by a Hubble radius $H^{-1}(t_1)=H_1$,
thereby becoming detached from the cosmological expansion and initiating collapse.
Within about a Hubble time, it will convert into a black hole (BH) of mass $M_{\rm PBH}=\xi M_1$.
Below, we assume that almost the entirety of the energy contained in the domain is deposited into the BH, so that $\xi\simeq 1$.

Since the interior of such a domain can exert repulsive gravity due to its substantial energy
dominance within the encompassing Hubble horizon, it may maintain a negative pressure,
particularly if the density contrast $\frac{\delta\rho}{\rho (t_1)}$ exceeds a threshold of around 10.
In such a scenario, the domain enters the Hubble radius at $t_1 \gg t_*$ and starts expanding faster than the background,
eventually reaching the inflationary vacuum and potentially developing a wormhole to a baby universe. Such a wormhole would
appear as a BH in the FRW Universe. In this paper, we focus on the regime with the most plausible collapse rather than expansion,
where $t_1 \approx t_*$, and thus, the size of the domain filled with $\Lambda_1$
is close to the Hubble radius at the moment when local dominance of the cosmological
term occurs, i.e., when the condition $\frac{\delta\rho}{\rho (t_*)} \approx 1$ is reached.

If the cosmological constant $\Lambda_1$ substantially exceeds its universe-averaged value $\Lambda_{\rm obs}$, then
the component $\rho_{\Lambda_{\rm obs}}(t_*)$ can be neglected in (\ref{densContr1}).
Therefore, the conditions for reaching a density contrast (\ref{densContr1}) exceeding unity can be described as
\begin{equation}
\label{eqdens}
\rho_{\Lambda_1}\equiv\frac{\Lambda_1}{8\pi G} \gtrsim \rho(t_1)\equiv \frac{3H^2_1}{8\pi G},
\end{equation}
where $G$ stands for the Newtonian constant, $H_1$ denotes the Hubble rate at $t_1$, and
the $\Lambda$ term is measured in units of the Planck mass squire. 

Thus, an overdense object with a size given by
\beq
\label{obj_size1}
l_1=H^{-1}_1=\sqrt{\frac{3}{\Lambda_1}}
\eeq
is formed, with its mass as measured by a distant observer being expressed as
\beq
\label{Mobs1}
M_1\simeq\frac{4\pi\rho{\Lambda_1}}{3H^3_1}=\frac{2\sqrt{3}}{\sqrt{\Lambda_1}G}.
\eeq
This mass is determined by the localized value of the cosmological constant $\Lambda_1$ within
a specific domain, which surpasses its universe-average value outside the domain.
The validity of Equation (\ref{Mobs1}) in the case of an overdense domain emerging due to
fluctuations in extra dimensions is rigorously proven in Appendix~\ref{App_C}.
Thus, it appears that the Schwarzschild radius of the above object, given by
\beq
\label{Rs1}
R_S=2GM_1= 4\sqrt{\frac{3}{\Lambda_1}},
\eeq
exceeds the size of the $\Lambda_1$ overdense domain given by Equation (\ref{obj_size1}).
Hence, assuming that the spherical shape of the domain is not significantly disturbed,
it will be converted into a BH. To account for the population of unevaporated black holes,
it is instructive to express the mass of such $\Lambda$-term-induced PBHs ($\Lambda$PBHs)
in units of solar masses:
\beq
\label{pbhMass1}
M_{\Lambda{\rm PBH}}=\frac{3.4\times 10^{-38}}{\sqrt{\Lambda/m^2_{\rm Pl}}}M_{\odot}.
\eeq

If we consider that the growth of supermassive black holes (SMBHs) observed
today originated from seed black holes, then this process must have commenced
in the early Universe, approximately 3 million years after the Big Bang,
with seeds heavier than $10^2\, M_{\odot}$. Additionally, there is evidence of
the existence of intermediate-mass black holes (IMBHs) with masses ranging from
$10^2\, M_{\odot}$ up to about $10^6\, M_{\odot}$. Thus, attributing the seeding
objects with masses from $10^2\, M_{\odot}$ to $10^6\, M_{\odot}$ to $\Lambda$PBHs
implies that they appeared as a result of the collapse of
domains containing $\Lambda$-terms spanning the range
\beq
\label{lambdaRange1}
\Lambda_1\simeq 10^{-78}m^2_{\rm Pl}\ \div \ 10^{-86}m^2_{\rm Pl}\ .
\eeq
In the subsequent analysis, we examine the comparability of the spectrum of the population
of $\Lambda$PBHs with the constraints on the abundance of PBHs within the considered mass range.

A domain of radius $R\approx\hinfinv$, filled with $\Lambda\neq\Lambda_{\rm obs}$, that emerges at the time moment $t_{\Lambda}$,
during inflation with a total duration $t_{\rm inf}$, when the Universe is
yet to inflate over $\Delta N_{\Lambda}=\hinf (t_{\rm inf}-t_{\Lambda})=N_{\rm inf}-N_{\Lambda}$ e-folds,
undergoes stretching during expansion as
\begin{equation}
\label{R1}
R(\Delta N_{\Lambda})\approx\hinfinv e^{\Delta N_{\Lambda}}.
\end{equation}
The number of domains created in a comoving volume $d{\bf V}$ within an e-fold interval $dN_{\Lambda}$ is determined by
\begin{equation}
\label{dN1}
dN=\Gamma_{\Lambda}\hinf^3e^{3N_{\Lambda}}d{\bf V}dN_{\Lambda},
\end{equation}
where $\Gamma_{\Lambda}$ represents the formation rate of domains with $\Lambda$ per Hubble
time-space volume $\hinf^{-4}$.
By expressing $N_{\Lambda}$ from (\ref{R1}), we can derive the number distribution of domains
with respect to their physical radius $R$ as
\begin{equation}
\label{dN2}
d{\cal N} = \Gamma_{\Lambda}\frac{e^{3N_{\Lambda}}d{\bf V}}{R^4}dR.
\end{equation}
Therefore, the number density in the physical inflationary volume $dV_{\rm inf}=e^{3N_{\Lambda}}d{\bf V}$ is
\begin{equation}
\label{n1}
\frac{dn}{dR}=\frac{d{\cal N}}{dRdV_{\rm inf}}=\frac{\Gamma_{\Lambda}}{R^4}.
\end{equation}
In the context of the setup discussed in this section, where domains of high-density contrast and obeying the condition (\ref{eqdens}) are considered, the distribution (\ref{n1}) covers a range of scales from $R_{\min} \simeq \hinfinv$ to $R_{\max} \equiv R(\Delta N_{\Lambda_1}) \approx \hinfinv e^{(N_{\rm inf}-N_{\Lambda_1})}$, where $N_{\Lambda_1}$ represents the number of e-folds when the probability of the appearance of at least one domain with $\Lambda_1$ becomes significant. 
This probability becomes notable over the course of the progression of inflation, which lasts for a sufficient number of $N_{\rm inf}$ e-folds necessary to address the horizon and flatness problems.

It is worth noting that if inflation were to occur above the TeV scale,
the comoving Hubble scale at the end of inflation would be less than one astronomical unit.
Consequently, a causally connected patch could
encompass our entire observable Universe today, which has a size
of about 30~Gpc, if there were more than 40 e-folds of inflation.
Similarly, if inflation occurred at the GUT scale ($\simeq$$10^{16}$~GeV),
then it would require more than 60 e-folds.
The upper bound on the value of the Hubble scale during slow-roll inflation provided by
Planck~\cite{Planck:Hinfl} is
\beq
\label{hlimit1}
\hinf= 6\times 10^{13}~{\rm GeV}\ .
\eeq

The mass distribution of black holes formed during the collapse of domains with values of $\Lambda$ deviating from $\Lambda_{\rm obs}$ is determined by the size distribution (\ref{n1}), scaled with respect to the expansion of the Universe (\ref{exp1}). This distribution can be expressed as
\begin{equation}
dn=\Gamma_{\Lambda}\frac{dR}{t_{eq}^{3/2}R^{5/2}},
\label{raspr}
\end{equation}
at the equality time $t_{\rm eq}=51$kyr. A convenient characteristic of this distribution, which facilitates
comparison of the PBH yield with constraints on their abundance in
different mass ranges (see, for instance, Figure 18
in~\cite{Caretal20}), is the mass density of PBHs
per logarithmic mass interval, expressed in units of the total density of the Universe:
\begin{equation}
\frac{d\Omega_{\rm PBH}}{d\ln M_{\rm PBH}}=\frac{1}{\rho_{eq}}\frac{dn}{d\ln M_{\rm PBH}}M_{\rm PBH},
\label{raspr1}
\end{equation}
where $\rho_{\rm eq}=m_{\rm Pl}^2/(6\pi t_{\rm eq}^2)$ represents the matter density at the time of equality.
Using (\ref{raspr}), we can obtain
\beq
\label{raspr2}
\frac{dn}{d\ln M_{\rm PBH}}=\frac{({4\pi})^{1/2}\Gamma_{\Lambda}}{3\sqrt{3}{t_{eq}^{3/2}}}\rho_{\Lambda}^{1/2}M_{\rm PBH}^{-1/2}\ ,
\eeq
where $\rho_{\Lambda}$ is the energy density contained in the domain filled with $\Lambda$ term which reads as
\beq
\label{rhoL1}
\rho_{\Lambda}=\frac{\Lambda m_{\rm Pl}^2}{8\pi}\ .
\eeq
where we recall that $\Lambda$ is expressed in units of $m_{\rm Pl}^2$.
Thus, (\ref{raspr1}) can be expanded as
\beq
\label{raspr3}
\frac{d\Omega_{\rm PBH}}{d\ln M_{\rm PBH}} = \sqrt{6}\pi \Gamma_{\Lambda}\left(\frac{\Lambda}
{m_{\rm Pl}^2}\right)^{1/2}t_{\rm eq}^{1/2}M_{\rm PBH}^{1/2}\approx 2.5\times 10^{66}
\Gamma_{\Lambda}\left(\frac{\Lambda}{m_{\rm Pl}^2}\right)^{1/2}\left(\frac{M_{\rm PBH}}{M_{\odot}}\right)^{1/2}\ .
\eeq
For those values of $\Lambda$ within the domains of inhomogeneities, as indicated in (\ref{lambdaRange1}),
the rate $\Gamma_{\Lambda}$ can be approximated as (a detailed derivation is provided in Appendix~\ref{App_B}):
\beq
\label{rate1}
\Gamma_{\Lambda}\simeq Q\Lambda\ ,
\eeq
where
\begin{equation}\label{Gamma2}
Q = \frac{8\pi^2}{3}\frac{m_P^2}{\hinf^4}\ .
\end{equation}
Substituting this into expression (\ref{raspr3}) and using relations (\ref{pbhMass1}), (\ref{rate1}), and
(\ref{Gamma2}),
we finally~obtain
\beq
\label{raspr4}
\frac{d\Omega_{\rm PBH}}{d\ln M_{\rm PBH}}\approx 2.6\times 10^{-26}
\left(\frac{\hinf}{m_{\rm Pl}}\right)^{-4}\left(\frac{M_{\rm PBH}}{M_{\odot}}\right)^{-1}\ .
\eeq
Comparing distribution (\ref{raspr4}) with the model
\beq
\label{contrPBH1}
\Omega_{\rm PBH}\sim10^9\beta\left(\frac{M_{\rm PBH}}{M_\odot}\right)^{-1/2}
\eeq
used in~\cite{Caretal20} to quote the constraints on the density fraction
$\beta$ deposited in PBHs at the moment of their formation, we arrive at the
following condition
\beq
\label{lambdaFromPBH_1}
\frac{\hinf}{m_{\rm Pl}}\simeq 6\times 10^{-9}\beta^{-1/4}\left(\frac{M_{\rm PBH}}{M_\odot}\right)^{-1/8}\ .
\eeq

Condition (\ref{lambdaFromPBH_1}) is useful for assessing the consistency of $\Lambda$PBH
formation with cosmological constraints on the abundance of PBHs across different mass ranges.~By analyzing the combined constraints on $\beta$ for a monochromatic mass function,
as presented in Figure~18 of~\cite{Caretal20}, we can verify the consistency of
considering $\Lambda$PBHs as candidates for seeding SMBHs and IMBHs,
taking into account the CMB constraints on the inflation scale.
For the seeding masses
$M_{\Lambda{\rm PBH}}\approx 10^2\, M_{\odot}$, the abundance is constrained to the level
of $\beta\approx 10^{-14}$~\cite{Caretal20}, which is saturated at the inflation energy scale
$\hinf\simeq 10^{-5}m_{\rm Pl}$.  At this level of precision, this can
be considered as the saturation point, ensuring that it does not exceed the CMB Planck limit (\ref{hlimit1}).
 Similar estimates of the inflation scale can be obtained for IMBHs
 in the mass range $10^2\, M_{\odot}\leq M_{\Lambda{\rm PBH}}\leq 5\times 10^5\, M_{\odot}$,
 where $\beta\approx 10^{-15}$~\cite{Caretal20}. For IMBH with masses $M_{\Lambda{\rm PBH}}\approx 10^6\, M_{\odot}$,
the energy scale $\hinf\simeq 10^{-7}m_{\rm Pl}$ saturates the constraint
$\beta\approx 10^{-7}$~\cite{Caretal20}. 
Similarly, the constraint $\beta\approx 3\times 10^{-7}$~\cite{Caretal20} imposed
for the mass scale $M_{\Lambda{\rm PBH}}\approx 10^{10}\, M_{\odot}$,
which is typically relevant for currently observed biggest SMBHs, is saturated at the inflation scale $\hinf\simeq 10^{-7}m_{\rm Pl}$.
Therefore, it can be concluded that $\Lambda$PBHs are cosmologically consistent for serving as seeds for SMBHs as well as explaining the observed IMBHs.

Additionally, $\Lambda$PBHs with masses as low as $M_{\Lambda{\rm PBH}}\approx 10^{-2}\, M_{\odot}$ remain compatible with the constraint
$\beta\simeq 10^{-11}$~\cite{Caretal20} imposed by the upper CMB inflation scale limit (\ref{hlimit1}).
This scenario corresponds to $\Lambda_1\simeq 10^{-76}m^2_{\rm Pl}$, where the formation mechanism of
$\Lambda$PBHs would generate about 10 times the mass of the Jupiter PBHs. Such PBHs could potentially account for a component of DM.

\section{Concluding Remarks}
\label{conclude}

It is theorized that the large-scale structure of the Universe was shaped by quantum fluctuations of scalar fields and/or metrics during inflation. These fluctuations, scaled exponentially with conserved amplitude, gave rise to primordial inhomogeneities, culminating in the formation of the cosmic web that represents the Universe's structure. In theories involving extra dimensions, fluctuations of fields within these dimensions can also be considered. However, the fate of these field fluctuations differs significantly from those associated with conventional four-dimensional scalar fields.

While conventional field fluctuations transform their energy density into radiation during the FRW stage through decay into other particle-like species, the energy density associated with scalar fields within the internal space remains unchanged, effectively stored within the scalar field itself, manifesting as a cosmological constant. Fluctuations of the scalar field within the internal space manifest as spatial variations of the local $\Lambda$ term. Domains containing an extraordinarily high cosmological constant may collapse into PBHs.

Upon investigating the mass distribution of such $\Lambda$PBHs, we find that it may satisfy existing cosmological constraints on the abundance of PBHs without conflicting with the upper bound on the inflation energy scale inferred from CMB measurements, within the mass range from $10^{-2}\, M_{\odot}$ up to $10^{10}\, M_{\odot}$. Of particular interest is the possibility of associating $\Lambda$PBHs with masses of $10^{2}\, M_{\odot}$ with seeds or supermassive black holes (SMBHs) and associating those with masses spanning the range from $10^{2}\, M_{\odot}$ to $10^{6}\, M_{\odot}$ with intermediate-mass black holes (IMBHs). The lightest $\Lambda$PBHs of masses $10^{-2}\, M_{\odot}$ can potentially contribute to the dark matter budget of the Universe.

The inevitable clustering of PBHs formed by the connected mechanism, driven by inflationary dynamics, leads to the formation of a Swiss cheese-like special structure of domains filled with high values of the $\Lambda$ term. This clustering may impact the characteristics of the observable spectrum of gravitational waves in the nanohertz frequency band, which are believed to be a signal from SMBHBs~\cite{clusterGW1}.

\section{Acknowledgments}

The work of S.G.R. was funded by the Ministry of Science and Higher Education of the Russian Federation, Project ``New Phenomena in Particle Physics and the Early Universe'' FSWU-2023-0073,
and the Kazan Federal University Strategic Academic Leadership Program. The work of A.A.P. was funded by the development program of the Volga Region Mathematical Center (agreement No. 075-02-2023-944).

\appendix
\section{Justification of Formula (\ref{Mobs1}) for Mass Measurement by a Distant Observer}
\label{App_C}

Here, we explore the conditions under which Equation~\eqref{Mobs1} can be reliably used.
We consider a three-dimensional space of volume $\Lambda_1 > \Lambda_0$. 
Our objective is to estimate the mass of such a region as observed by a distant observer.
For our estimation, we make several assumptions: the field distribution varies slowly,
allowing us to neglect its time dependence; we operate far below the inflationary scale,
implying that the term $aR^2$ is negligible; the domain with the higher energy density has
approximately spherical geometry; and the amount of ordinary matter is negligible.

We set the four-dimensional effective action Equation~(\ref{S_eff}) as 
 \begin{eqnarray} \label{S4}
S_4 & = & \frac{m_P ^{2}}{2}  \int  dt dv d\theta d\varphi \sqrt{|g_{4}|} \, [aR^2 + R -2\Lambda(v]\
\end{eqnarray}
and assume that the D dimensional metric depends on the radial coordinate $v$, leading to the interval in the form
\beq \label{31}
ds^2 =A(v)d t^2 -\frac{1}{B(v)}d v^2- v^2d\Omega_2^2,
\eeq
which evolves the action (\ref{S4}) into the following expression
\beq \label{Ein}
S_4  = \frac{m_P ^{2}}{2}  \int  dt dv d\theta d\varphi \sqrt{|g_{4}|} \, [ R -2\Lambda(v)]\ ,
\eeq
where the term $aR^2$ is neglected.
Now, the situation is essentially simplified, allowing for the analytical evaluation of the mass using the nontrivial equations of the theory
\beq
\label{tt9}
\frac{B'}{v} +\frac{B}{v^2} -\frac{1}{v^2} +\Lambda =0,
\eeq
\beq \label{rr9}
\frac{B}{v} \frac{A'}{A} +\frac{B}{v^2} -\frac{1}{v^2} +\Lambda   =0,
\eeq
\beq  \label{3thth}
\frac{B}{2} \frac{A''}{A}  -\frac{B}{4} \frac{{A'}^2}{A^2} +\frac{B'}{4} \frac{A'}{A} +\frac{B}{2v} \frac{A'}{A}  +\frac{B'}{2 v}  +\Lambda   =0.
\eeq
Subtracting the first two equations yields
\begin{equation} \label{BA}
    B(v)=A(v)\ ,
\end{equation}
so that Equations (\ref{tt9}) and (\ref{rr9}) are reduced to a single equation
\beq
\label{singleEQ1}
\frac{A'}{v} +\frac{A}{v^2} -\frac{1}{v^2} +\Lambda   =0,
\eeq
with the solution
\beq\label{Av}
A(v) = \frac{1}{v} \int_0^v \Big( 1 -\Lambda(v) v^2 \Big) dv -\frac{2 C_1}{v}\ .
\eeq
We assume that the value of $\Lambda$ remains
constant within the sphere of radius $l_*$, denoted as $\Lambda(l_*) = \Lambda_1$.
Therefore, inside the sphere, (\ref{Av}) is converted into the expression
\beq \label{Ain}
A(v)\vert_{v\leq l_*} =1 -\frac{\Lambda_1 l_*^2}{3}-\frac{2 C_1}{l_*}\ .
\eeq
Since, in our setup, there is no point-like mass producing the singularity in the center,
$C_1=0$, and hence, we obtain the well-known de Sitter metric.
For a distant observer in the Minkowski space we assume $\Lambda(v)=0$ everywhere outside the sphere of radius $l_*$ so that one can derive
\beq \label{Aout}
A(v)\vert_{v>l_*} = A(v)\vert_{v\leq l_*} +\frac{1}{v} \int_{l_*}^v  dv =1 - \frac{\Lambda_1 l_*^3}{ 3  v}.
\eeq
Thus, the mass of the domain filled with
$\Lambda_1$ reads $2 M_{\infty} = {\Lambda_1 l_*^3 m_P^2}/{ 3}$ and finally can be expressed as
\begin{equation}\label{M}
    M_{\infty} = \frac{4 \pi \rho_\Lambda l_*^3}{ 3}
\end{equation}
providing the definition $\Lambda_1 = 8 \pi \rho_\Lambda /m_P^2$.
Note that taking into account \eqref{BA}, \eqref{Ain}, and \eqref{Aout}, Equation \eqref{3thth} becomes an identity.

\section{Probability of Domain Formation with Specific \\ Energy Density during~Inflation}
\label{App_B}

To estimate the volume fraction $dP(\Lambda)$ filled with a specific value of the energy density $\rho_{\Lambda}=\Lambda m_P^2/8\pi$,
we need to relate the field fluctuations in the extra dimensions during the inflationary stage to the parameter $\Lambda$.
Following the approach outlined in~\cite{2021arXiv210908373R}, we consider the extra metric as the background one,
allowing the parameter $c_{\text{eff}}$ to vary together with the scalar field $\zeta$.
The scalar field fluctuates intensively during inflation. 

Therefore, our first approximation is to assume a pure de Sitter metric, such that the function $\zeta(x,y)$ is governed by the equation
\begin{equation}
\zeta'' + \left(4 \gamma' + (n-1) \frac{r'}{r}\right) \zeta' - V_{\zeta} =0.
\end{equation}
This equation, one of Equation set \eqref{EE}, has an asymptote of $\zeta_1(u)$ in a chosen 3D
volume under the horizon. 
The surrounding 3D space is characterized by another static configuration $\zeta_0(u)\neq \zeta_1(u)$.

The parameter value $c_{\text{eff}}[\zeta_0]$ expressed by the last equation in Equation~\eqref{mPl}, is assumed to be small in order to avoid disrupting the dynamics of the inflationary stage. Additionally, it is assumed to tend to a post-inflationary value of $10^{-123}m_P^2$. Therefore, the quantity
\begin{equation}\label{dceff}
c_{\text{eff}}[\zeta_1] \neq 0,
\end{equation}
is responsible for the excess energy density.

The scalar field action \eqref{SD} can be reduced to the standard form
\begin{equation}\label{Ss}
S_{\text{scalar}}=\frac{1}{2}\int d^4x \sqrt{g_4} \int d^n y \sqrt{g_n} [(\partial \phi)^2 - m^2\phi^2]
\end{equation}
by using the substitution
\begin{equation}\label{pz}
\phi=m^{\frac{D-2}{2}}\zeta\ .
\end{equation}

Exact calculation of the probability for a specific fluctuation $\zeta_1(u)$ is quite difficult. 
For estimation purposes, we can use an approximation in the spirit of the Kaluza--Klein approach, where the scalar field is represented in the form
\begin{equation}\label{sdec3}
\phi_1(x,y)=\phi_0(y)+ \delta\phi(x,y), \quad \delta\phi(x,y)= \sum_{a} \phi^{(a)}(x) Y_a (y),
\end{equation}
with the standard normalization
\begin{equation}
1=\int d^n y \sqrt{g_n} Y_a(y)^2,
\end{equation}
where $\phi_0(y)$ is an initial static classical part of the scalar field for which $c_{\text{eff}}=0$, and the difference $\delta\phi(x,y)$ is decomposed into a series of orthogonal normalized functions $Y_a$. Here, for convenience, we use the dimensionalities $[\zeta]=1, [\phi]=m^{1+n/2}, [Y_a]=m^{n/2}\to [\phi^{(a)}]=m$ and neglect the internal $n$-dimensional metric variation.

Substituting \eqref{sdec3} into \eqref{Ss}, we obtain the action in the form
\begin{equation}\label{Ssfin}
S=\sum_a \frac {1}{2}\int d^{4}x\sqrt{g_{4}}[\partial_{\mu}\phi^{(a)}\partial^{\mu}\phi^{(a)} - \mu_a^2\phi^{(a)2}] + S[\zeta_0], \quad \mu_a^2=m^2+\lambda_a\ ,
\end{equation}
at the inflationary stage, where the term $S[\zeta_0]$ tends to zero at the present time by definition. The discrete set of eigenvalues $\lambda_a,, a=0,\pm 1,\pm 2...$ depends on the specific form of the extra space metric.
Assuming that $\phi^{(a)}$ represents long-wave fluctuations that freeze at the inflationary stage, i.e., $\phi^{(a)}=\text{const}$,
we obtain the relation between the Lambda term and the scalar field fluctuation under the horizon stems from the equalities
\begin{equation}\label{ceffL}
-2\Lambda \equiv c_{\text{eff}}[\zeta(u)]= - \frac{\sum_a\mu_a^2 \phi^{(a)2}}{m_P^2},
\end{equation}
as follows from \eqref{S_eff} and \eqref{Ssfin}.


The excitation amplitudes $\phi^{(a)}$ act as independent free fields with an initial
amplitude equal to $\phi_0(u)$~\cite{2021arXiv210908373R}. The probability of finding a set of functions $\phi^{(a)}$
\begin{equation}\label{Probt}
dP(\{\phi_a\})\simeq \left[\prod_a d\phi^{(a)} \cdot\sqrt{q_a/\pi}\right]\exp\left[-\sum_a q_a\phi^{(a)2}\right],\quad q_a = \frac{4\pi^2}{3}\frac{\mu_a^2}{H^4}.
\end{equation}

According to \eqref{ceffL},
\begin{equation}
\sum_a q_a\phi^{(a)2}=\frac{4\pi^2}{3}H^{-4}\sum_a \mu_a^2\phi^{(a)2}=\frac{8 \pi^2}{3}\frac{m_P^2}{H^4}\Lambda
\end{equation}
The final expression for the probability is
\begin{equation}\label{ProbL}
dP(\Lambda)\simeq {d\Lambda}\cdot {Q} \exp \left[-Q\Lambda\right],\quad Q = \frac{8\pi^2}{3}\frac{m_P^2}{H^4},
\end{equation}
where the pre-exponent follows from the normalization.~The number of domains with specific $\Lambda$ is related to the probability as
\begin{equation}\label{dn}
    dn=e^{3N}dP(\Lambda).
\end{equation}


\begin{thebibliography}{999}

\bibitem{1967SvA....10..602Z}
Zel'dovich, Y.B.; Novikov, I.D.
\newblock {The Hypothesis of Cores Retarded during Expansion and the Hot
  Cosmological Model}.
\newblock {\em Sov. Astron. J.} {\bf 1967}, {\em 10},~602--603.

\bibitem{1971MNRAS.152...75H}
Hawking, S.
\newblock {Gravitationally collapsed objects of very low mass}.
\newblock {\em Mon. Not. R. Astron. Soc.} {\bf 1971}, {\em 152},~75.

\bibitem{1974MNRAS.168..399C}
{Carr}, B.J.; {Hawking}, S.W.
\newblock {Black holes in the early Universe}.
\newblock {\em Mon. Not. Roy. Astron. Soc.} {\bf 1974}, {\em 168},~399--416.
\newblock {\url{https://doi.org/10.1093/mnras/168.2.399}}.

\bibitem{1980PhLB...97..383K}
Khlopov, M.Y.; Polnarev, A.G.
\newblock {Primordial Black Holes As A Cosmological Test Of Grand Unification}.
\newblock {\em Phys. Lett.~{\bf{B}}} {\bf 1980}, {\em 97},~383--387.
\newblock {\url{https://doi.org/10.1016/0370-2693(80)90624-3}}.

\bibitem{1985UsFiN.145..369P}
{Polnarev}, A.G.; {Khlopov}, M.I.
\newblock {Cosmology, primary black holes and supermassive particles}.
\newblock {\em Uspekhi Fizicheskikh Nauk} {\bf 1985}, {\em 145},~369--401.

\bibitem{2010RAA....10..495K}
{Khlopov}, M.Y.
\newblock {Primordial black holes}.
\newblock {\em Res. Astron. Astrophys.} {\bf 2010}, {\em
  10},~495--528.
\newblock {\url{https://doi.org/10.1088/1674-4527/10/6/001}}.

\bibitem{2019EPJC...79..246B}
{Belotsky}, K.M.; {Dokuchaev}, V.I.; {Eroshenko}, Y.N.; {Esipova}, E.A.;
  {Khlopov}, M.Y.; {Khromykh}, L.A.; {Kirillov}, A.A.; {Nikulin}, V.V.;
  {Rubin}, S.G.; {Svadkovsky}, I.V.
\newblock {Clusters of Primordial Black Holes}.
\newblock {\em Eur. Phys. J. C} {\bf 2019}, {\em 79},~246.
\newblock {\url{https://doi.org/10.1140/epjc/s10052-019-6741-4}}.

\bibitem{2018CQGra..35f3001S}
{Sasaki}, M.; {Suyama}, T.; {Tanaka}, T.; {Yokoyama}, S.
\newblock {Primordial black
 holes{\textemdash}perspectives in gravitational
  wave astronomy}.
\newblock {\em Class. Quantum Gravity} {\bf 2018}, {\em 35},~063001.
\newblock {\url{https://doi.org/10.1088/1361-6382/aaa7b4}}.

\bibitem{2021PhRvD.103h3518G}
{Gundhi}, A.; {Ketov}, S.V.; {Steinwachs}, C.F.
\newblock {Primordial black hole dark matter in dilaton-extended two-field
  Starobinsky inflation}.
\newblock {\em Phys. Rev.~{\bf{D}}} {\bf 2021}, {\em 103},~083518.
\newblock {\url{https://doi.org/10.1103/PhysRevD.103.083518}}.

\bibitem{Caretal20}
Carr, B.; Kohri, K.; Sendouda, Y.; Yokoyama, J.
\newblock {Constraints on primordial black holes}.
\newblock {\em Rept. Prog. Phys.} {\bf 2021}, {\em 84},~116902.
\newblock {\url{https://doi.org/10.1088/1361-6633/ac1e31}}.

\bibitem{pbhMech2}
Raidal, M.; Vaskonen, V.; Veerm\"ae, H.
\newblock {Gravitational Waves from Primordial Black Hole Mergers}.
\newblock {\em  J. Cosmol. Astropart. Phys.} {\bf 2017}, {\em 09},~037.
\newblock {\url{https://doi.org/10.1088/1475-7516/2017/09/037}}.

\bibitem{pbhMech3}
Raidal, M.; Spethmann, C.; Vaskonen, V.; Veerm\"ae, H.
\newblock {Formation and Evolution of Primordial Black Hole Binaries in the
  Early Universe}.
\newblock {\em J. Cosmol. Astropart. Phys.} {\bf 2019}, {\em 02},~018.
\newblock {\url{https://doi.org/10.1088/1475-7516/2019/02/018}}.

\bibitem{LVK1}
Abbott, B.P.  et~al. [LIGO Scientific Collaboration and Virgo Collaboration]
\newblock {Binary Black Hole Mergers in the first Advanced LIGO Observing Run}.
\newblock {\em Phys. Rev. X} {\bf 2016}, {\em 6},~041015.
\newblock Erratum in \emph{Phys. Rev. X} \textbf{2018}, \emph{8}, 039903.
  {\url{https://doi.org/10.1103/PhysRevX.6.041015}}.

\bibitem{LVK2}
Abbott, R.  et~al. [LIGO Scientific Collaboration and Virgo Collaboration].
\newblock {GWTC-2: Compact Binary Coalescences Observed by LIGO and Virgo
  During the First Half of the Third Observing Run}.
\newblock {\em Phys. Rev. X} {\bf 2021}, {\em 11},~021053.
\newblock {\url{https://doi.org/10.1103/PhysRevX.11.021053}}.

\bibitem{LVKdetails1}
Sasaki, M.; Suyama, T.; Tanaka, T.; Yokoyama, S.
\newblock {Primordial Black Hole Scenario for the Gravitational-Wave Event
  GW150914}.
\newblock {\em Phys. Rev. Lett.} {\bf 2016}, {\em 117},~061101.
\newblock Erratum in \emph{Phys. Rev. Lett.} \textbf{2018}, \emph{121}, 059901.
  {\url{https://doi.org/10.1103/PhysRevLett.117.061101}}.

\bibitem{LVKdetails2}
Hall, A.; Gow, A.D.; Byrnes, C.T.
\newblock {Bayesian analysis of LIGO-Virgo mergers: Primordial vs.
  astrophysical black hole populations}.
\newblock {\em Phys. Rev. D} {\bf 2020}, {\em 102},~123524.
\newblock {\url{https://doi.org/10.1103/PhysRevD.102.123524}}.

\bibitem{LVKdetails3}
Jedamzik, K.
\newblock {Primordial Black Hole Dark Matter and the LIGO/Virgo observations}.
\newblock {\em J. Cosmol. Astropart. Phys.} {\bf 2020}, {\em 09},~022.
\newblock {\url{https://doi.org/10.1088/1475-7516/2020/09/022}}.

\bibitem{LVKdetails4}
De~Luca, V.; Franciolini, G.; Pani, P.; Riotto, A.
\newblock {Primordial Black Holes Confront LIGO/Virgo data: Current situation}.
\newblock {\em J. Cosmol. Astropart. Phys.} {\bf 2020}, {\em 06},~044.
\newblock {\url{https://doi.org/10.1088/1475-7516/2020/06/044}}.

\bibitem{LVKdetails5}
Franciolini, G.; Baibhav, V.; De~Luca, V.; Ng, K.K.Y.; Wong, K.W.K.; Berti, E.;
  Pani, P.; Riotto, A.; Vitale, S.
\newblock {Searching for a subpopulation of primordial black holes in
  LIGO-Virgo gravitational-wave data}.
\newblock {\em Phys. Rev. D} {\bf 2022}, {\em 105},~083526.
\newblock {\url{https://doi.org/10.1103/PhysRevD.105.083526}}.

\bibitem{LVKdetails6}
Romero-Rodriguez, A.; Martinez, M.; Pujol\`as, O.; Sakellariadou, M.; Vaskonen,
  V.
\newblock {Search for a Scalar Induced Stochastic Gravitational Wave Background
  in the Third LIGO-Virgo Observing Run}.
\newblock {\em Phys. Rev. Lett.} {\bf 2022}, {\em 128},~051301.
\newblock {\url{https://doi.org/10.1103/PhysRevLett.128.051301}}.

\bibitem{smbh1}
Ferrarese, L.; Ford, H.
\newblock {Supermassive black holes in galactic nuclei: Past, present and
  future research}.
\newblock {\em Space Sci. Rev.} {\bf 2005}, {\em 116},~523--624.
\newblock {\url{https://doi.org/10.1007/s11214-005-3947-6}}.

\bibitem{smbh2}
Gultekin, K.; Richstone, D.O.; Gebhardt, K.; Lauer, T.R.; Tremaine, S.; Aller, M.C.; Bender, R.; Dressler, A.; Faber, S.M.; Filippenko, A.V.;  et~al.
\newblock {The M-sigma and M-L Relations in Galactic Bulges and Determinations
  of their Intrinsic Scatter}.
\newblock {\em Astrophys. J.} {\bf 2009}, {\em 698},~198--221.
\newblock {\url{https://doi.org/10.1088/0004-637X/698/1/198}}.

\bibitem{smbh3}
Kormendy, J.; Ho, L.C.
\newblock {Coevolution (Or Not) of Supermassive Black Holes and Host Galaxies}.
\newblock {\em Ann. Rev. Astron. Astrophys.} {\bf 2013}, {\em 51},~511--653.
\newblock {\url{https://doi.org/10.1146/annurev-astro-082708-101811}}.

\bibitem{acr1}
Ali-Ha\"{i}moud, Y.; Kamionkowski, M.
\newblock {Cosmic microwave background limits on accreting primordial black
  holes}.
\newblock {\em Phys. Rev. D} {\bf 2017}, {\em 95},~043534.
\newblock {\url{https://doi.org/10.1103/PhysRevD.95.043534}}.

\bibitem{acr2}
Bean, R.; Magueijo, J.
\newblock {Could supermassive black holes be quintessential primordial black
  holes?}
\newblock {\em Phys. Rev. D} {\bf 2002}, {\em 66},~063505.
\newblock {\url{https://doi.org/10.1103/PhysRevD.66.063505}}.

\bibitem{acr3}
Kawasaki, M.; Kusenko, A.; Yanagida, T.T.
\newblock {Primordial seeds of supermassive black holes}.
\newblock {\em Phys. Lett. B} {\bf 2012}, {\em 711},~1--5.
\newblock {\url{https://doi.org/10.1016/j.physletb.2012.03.056}}.

\bibitem{acr4}
Clesse, S.; Garc\'{i}a-Bellido, J.
\newblock {The clustering of massive Primordial Black Holes as Dark Matter:
  measuring their mass distribution with Advanced LIGO}.
\newblock {\em Phys. Dark Univ.} {\bf 2017}, {\em 15},~142--147.
\newblock {\url{https://doi.org/10.1016/j.dark.2016.10.002}}.

\bibitem{acr5}
Clesse, S.; Garc\'{i}a-Bellido, J.
\newblock {Seven Hints for Primordial Black Hole Dark Matter}.
\newblock {\em Phys. Dark Univ.} {\bf 2018}, {\em 22},~137--146.
\newblock {\url{https://doi.org/10.1016/j.dark.2018.08.004}}.

\bibitem{acr6}
Serpico, P.D.; Poulin, V.; Inman, D.; Kohri, K.
\newblock {Cosmic microwave background bounds on primordial black holes
  including dark matter halo accretion}.
\newblock {\em Phys. Rev. Res.} {\bf 2020}, {\em 2},~023204.
\newblock {\url{https://doi.org/10.1103/PhysRevResearch.2.023204}}.

\bibitem{fluc1}
Nakama, T.; Suyama, T.; Yokoyama, J.
\newblock {Supermassive black holes formed by direct collapse of inflationary
  perturbations}.
\newblock {\em Phys. Rev. D} {\bf 2016}, {\em 94},~103522.
\newblock {\url{https://doi.org/10.1103/PhysRevD.94.103522}}.

\bibitem{fluc2}
Nakama, T.; Carr, B.; Silk, J.
\newblock {Limits on primordial black holes from $\mu$ distortions in cosmic
  microwave background}.
\newblock {\em Phys. Rev. D} {\bf 2018}, {\em 97},~043525.
\newblock {\url{https://doi.org/10.1103/PhysRevD.97.043525}}.

\bibitem{galform1}
Carr, B.; Silk, J.
\newblock {Primordial Black Holes as Generators of Cosmic Structures}.
\newblock {\em Mon. Not. Roy. Astron. Soc.} {\bf 2018}, {\em 478},~3756--3775.
\newblock {\url{https://doi.org/10.1093/mnras/sty1204}}.

\bibitem{galform2}
Inman, D.; Ali-Ha\"{i}moud, Y.
\newblock {Early structure formation in primordial black hole cosmologies}.
\newblock {\em Phys. Rev. D} {\bf 2019}, {\em 100},~083528.
\newblock {\url{https://doi.org/10.1103/PhysRevD.100.083528}}.

\bibitem{IMBH_1}
Greene, J.E.; Strader, J.; Ho, L.C.
\newblock {Intermediate-Mass Black Holes}.
\newblock {\em Ann. Rev. Astron. Astrophys.} {\bf 2020}, {\em 58},~257--312.
\newblock {\url{https://doi.org/10.1146/annurev-astro-032620-021835}}.

\bibitem{smbhdm1}
Carr, B.; Kuhnel, F.; Visinelli, L.
\newblock {Constraints on Stupendously Large Black Holes}.
\newblock {\em Mon. Not. Roy. Astron. Soc.} {\bf 2021}, {\em 501},~2029--2043,
  \href{http://arxiv.org/abs/2008.08077}{{\normalfont
  [arXiv:astro-ph.CO/2008.08077]}}.
\newblock {\url{https://doi.org/10.1093/mnras/staa3651}}.

\bibitem{jwst1}
Liu, B.; Bromm, V.
\newblock {Accelerating Early Massive Galaxy Formation with Primordial Black
  Holes}.
\newblock {\em Astrophys. J. Lett.} {\bf 2022}, {\em 937},~L30,
  \href{http://arxiv.org/abs/2208.13178}{{\normalfont
  [arXiv:astro-ph.CO/2208.13178]}}.
\newblock {\url{https://doi.org/10.3847/2041-8213/ac927f}}.

\bibitem{pta1}
Goncharov, B.; Shannon, R.M.; Reardon, D.J.; Hobbs, G.; Zic, A.; Bailes, M.; Curyło, M.; Dai, S.; Kerr, M.; Lower, M.E.;  et~al.
\newblock {On the Evidence for a Common---Spectrum Process in the Search for
  the Nanohertz Gravitational---Wave Background with the Parkes Pulsar Timing
  Array}.
\newblock {\em Astrophys. J. Lett.} {\bf 2021}, {\em 917},~L19.
\newblock {\url{https://doi.org/10.3847/2041-8213/ac17f4}}.

\bibitem{pta2}
Chen, S.; Caballero, R.N.; Guo, Y.J.; Chalumeau, A.; Liu, K.; Shaifullah, G.;  Lee, K.J.; Babak, S.; Desvignes, G.; Parthasarathy, A.; et~al.
\newblock {Common-red-signal analysis with 24-yr high-precision timing of the
  European Pulsar Timing Array: inferences in the stochastic gravitational-wave
  background search}.
\newblock {\em Mon. Not. Roy. Astron. Soc.} {\bf 2021}, {\em 508},~4970--4993.
\newblock {\url{https://doi.org/10.1093/mnras/stab2833}}.

\bibitem{pta3}
Antoniadis, J.; Arumugam, P.; Arumugam, S.; Babak, S.; Bagchi, M.; Nielsen, A.S.B.; Bassa, C.G.; Bathula, A.; Berthereau, A.; Bonetti, M.;  et~al.
\newblock {The second data release from the European Pulsar Timing Array---III.
  Search for gravitational wave signals}.
\newblock {\em Astron. Astrophys.} {\bf 2023}, {\em 678},~A50.
\newblock {\url{https://doi.org/10.1051/0004-6361/202346844}}.

\bibitem{pta4}
Reardon, D.J.; Zic, A.; Shannon, R.M.; Hobbs, G.B.; Bailes, M.; Di Marco, V.;  AKapur, g.; Rogers, A.F.; Thrane, E.; Askew, J.; et~al.
\newblock {Search for an Isotropic Gravitational-wave Background with the
  Parkes Pulsar Timing Array}.
\newblock {\em Astrophys. J. Lett.} {\bf 2023}, {\em 951},~L6.
\newblock {\url{https://doi.org/10.3847/2041-8213/acdd02}}.

\bibitem{pta5}
Xu, H.; Chen, S.; Guo, Y.; Jiang, J.; Wang, B.; Xu, J.; Xue, Z.; Caballero, R.N.; Yuan, J.; Xu, Y.;  et~al.
\newblock {Searching for the Nano-Hertz Stochastic Gravitational Wave
  Background with the Chinese Pulsar Timing Array Data Release I}.
\newblock {\em Res. Astron. Astrophys.} {\bf 2023}, {\em 23},~075024.
\newblock {\url{https://doi.org/10.1088/1674-4527/acdfa5}}.

\bibitem{pta6}
Agazie, G.; Anumarlapudi, A.; Archibald, A.M.; Baker, P.T.; Bécsy, B.; Blecha, L.; Bonilla, A.; Brazier, A.; Brook, P.R.; Burke-Spolaor, S.;  et~al.
\newblock {The NANOGrav 15 yr Data Set: Constraints on Supermassive Black Hole
  Binaries from the Gravitational-wave Background}.
\newblock {\em Astrophys. J. Lett.} {\bf 2023}, {\em 952},~L37.
\newblock {\url{https://doi.org/10.3847/2041-8213/ace18b}}.

\bibitem{pta7}
Agazie, G.;  Anumarlapudi, A.; Archibald, A.M.; Arzoumanian, Z.; Baker, P.T.; Bécsy, B.; Blecha, L.; Brazier, A.; Brook, P.R.; Burke-Spolaor, S.;  et~al.
\newblock {The NANOGrav 15 yr Data Set: Evidence for a Gravitational-wave
  Background}.
\newblock {\em Astrophys. J. Lett.} {\bf 2023}, {\em 951},~L8.
\newblock {\url{https://doi.org/10.3847/2041-8213/acdac6}}.

\bibitem{Valli:2024nbj}
Valli, R.; Tiede, C.; Vigna-G\'omez, A.; Cuadra, J.; Siwek, M.; Ma, J.Z.;
  D'Orazio, D.J.; Zrake, J.; de~Mink, S.E.
\newblock {Long-term Evolution of Binary Orbits Induced by Circumbinary Disks}. \emph{arXiv}
  {\bf 2024}.
  \url{http://arxiv.org/abs/2401.17355}.

\bibitem{Alonso-Alvarez:2024gdz}
Alonso-\'Alvarez, G.; Cline, J.M.; Dewar, C.
\newblock {Self-interacting dark matter solves the final parsec problem of
  supermassive black hole mergers}. \emph{arXiv} {\bf 2024}.
  \url{http://arxiv.org/abs/2401.14450}

\bibitem{Buchmuller:2024zzk}
Buchmuller, W.
\newblock {Metastable strings and grand unification}. \emph{arXiv}
\newblock   \textbf{2024}.  \url{http://arxiv.org/abs/2401.13333}.

\bibitem{Winkler:2024olr}
Winkler, M.W.; Freese, K.
\newblock {Origin of the Stochastic Gravitational Wave Background: First-Order
  Phase Transition vs. Black Hole Mergers}. \emph{arXiv} {\bf 2024}.
\url{http://arxiv.org/abs/2401.13729}.

\bibitem{Conaci:2024tlc}
Conaci, A.; Delle~Rose, L.; Dev, P.S.B.; Ghoshal, A.
\newblock {Slaying Axion-Like Particles via Gravitational Waves and Primordial
  Black Holes from Supercooled Phase Transition}. \emph{arXiv} {\bf 2024}.
\url{http://arxiv.org/abs/2401.09411}.

\bibitem{Choudhury:2024one}
Choudhury, S.; Karde, A.; Panda, S.; Sami, M.
\newblock {Realisation of the ultra-slow roll phase in Galileon inflation and
  PBH overproduction}. \emph{arXiv} {\bf 2024}.
\url{http://arxiv.org/abs/2401.10925}.

\bibitem{Padmanabhan:2024nvv}
Padmanabhan, H.; Loeb, A.
\newblock {Constraints on Supermassive Black Hole Binaries from JWST and
  NANOGrav}. \emph{arXiv} {\bf 2024}.
\url{http://arxiv.org/abs/2401.04161}.

\bibitem{Hu:2023oiu}
Hu, L.; Cai, R.G.; Wang, S.J.
\newblock {Distinctive GWBs from eccentric inspiraling SMBH binaries with a DM
  spike}. \emph{arXiv} {\bf 2023}.
\url{http://arxiv.org/abs/2312.14041}.

\bibitem{Lacy:2023kbb}
Lacy, M.; Engholm, A.; Farrah, D.; Ejercito, K.
\newblock {Constraints on Cosmological Coupling from the Accretion History of
  Supermassive Black Holes}.
\newblock {\em Astrophys. J. Lett.} {\bf 2024}, {\em 961},~L33.
\newblock {\url{https://doi.org/10.3847/2041-8213/ad1b5f}}.

\bibitem{Eichhorn:2023iab}
Eichhorn, A.; Fernandes, P.G.S.; Held, A.; Silva, H.O.
\newblock {Breaking black-hole uniqueness at supermassive scales}. \emph{arXiv} {\bf 2023}.
\url{http://arxiv.org/abs/2312.11430}.

\bibitem{Zhang:2023jrk}
Zhang, F.
\newblock {Final parsec evolution in the presence of intermediate mass black
  holes}. \emph{arXiv} {\bf 2023}.
\url{http://arxiv.org/abs/2312.11847}.

\bibitem{Sato-Polito:2023gym}
Sato-Polito, G.; Zaldarriaga, M.; Quataert, E.
\newblock {Where are NANOGrav's big black holes?}. \emph{arXiv} {\bf 2023}.
\url{http://arxiv.org/abs/2312.06756}.

\bibitem{Liu:2023pvq}
Liu, B.; Bromm, V.
\newblock {Impact of primordial black holes on the formation of the first stars
  and galaxies}. \emph{arXiv} {\bf 2023}.
\url{http://arxiv.org/abs/2312.04085}.

\bibitem{Ellis:2023iyb}
Ellis, J.; Fairbairn, M.; Urrutia, J.; Vaskonen, V.
\newblock {Probing supermassive black hole seed scenarios with gravitational
  wave measurements}. \emph{arXiv} {\bf 2023}.
\url{http://arxiv.org/abs/2312.02983}.

\bibitem{Huang:2023klk}
Huang, H.L.; Jiang, J.Q.; Piao, Y.S.
\newblock {Merger rate of supermassive primordial black hole binaries}. \emph{arXiv} {\bf
  2023}.
\url{http://arxiv.org/abs/2312.00338}.

\bibitem{Bromley:2023yfi}
Bromley, B.C.; Sandick, P.; Shams Es~Haghi, B.
\newblock {Supermassive Black Hole Binaries in Ultralight Dark Matter}. \emph{arXiv} {\bf
  2023}.
\url{http://arxiv.org/abs/2311.18013}.

\bibitem{Davis:2023vyy}
Davis, M.C.; Grace, K.E.; Trump, J.R.; Runnoe, J.C.; Henkel, A.; Blecha, L.;
  Brandt, W.N.; Casey-Clyde, J.A.; Charisi, M.; Witt, C.
\newblock {Reliable Identification of Binary Supermassive Black Holes from
  Rubin Observatory Time-Domain Monitoring}. \emph{arXiv} {\bf 2023}.
\url{http://arxiv.org/abs/2311.10851}.

\bibitem{Harris:2023xab}
Harris, C.; Gultekin, K.
\newblock {Connecting Core Galaxy Properties to the Massive Black Hole Binary
  Population}. \emph{arXiv} {\bf 2023}.
\url{http://arxiv.org/abs/2311.04877}.

\bibitem{Koo:2023gfm}
Koo, H.; Bak, D.; Park, I.; Hong, S.E.; Lee, J.W.
\newblock {Final parsec problem of black hole mergers and ultralight dark
  matter}. \emph{arXiv} {\bf 2023}.
\url{http://arxiv.org/abs/2311.03412}.

\bibitem{Ramazanov:2023eau}
Ramazanov, S.
\newblock {Spectrum of gravitational waves from long-lasting primordial
  sources}. \emph{arXiv} {\bf 2023}.
\url{http://arxiv.org/abs/2310.19148}.

\bibitem{DOrazio:2023rvl}
D'Orazio, D.J.; Charisi, M.
\newblock {Observational Signatures of Supermassive Black Hole Binaries}.
\newblock \emph{arXiv}   \textbf{2023}.  \url{http://arxiv.org/abs/2310.16896}.

\bibitem{Kasai:2023qic}
Kasai, K.; Kawasaki, M.; Kitajima, N.; Murai, K.; Neda, S.; Takahashi, F.
\newblock {Primordial Origin of Supermassive Black Holes from Axion Bubbles}. \emph{arXiv}
  {\bf 2023}.
\url{http://arxiv.org/abs/2310.13333}.

\bibitem{Stamou:2023vft}
Stamou, I.; Clesse, S.
\newblock {Primordial Black Holes without fine-tuning from a light stochastic
  spectator field}. \emph{arXiv} {\bf 2023}.
\url{http://arxiv.org/abs/2310.04174}.

\bibitem{Dolgov:2023ijt}
Dolgov, A.D.
\newblock {Tension between HST/JWST and $\Lambda$CDM Cosmology, PBH, and
  Antimatter in the Galaxy}.
\newblock In Proceedings of the {14th Frascati workshop on Multifrequency
  Behaviour of High Energy Cosmic Sources}, Palermo, Italy, 12--17 June 2023.

\bibitem{Huang:2023nes}
Huang, F.; Bi, Y.C.; Cao, Z.; Huang, Q.G.
\newblock {Impacts of Gravitational-Wave Background from Supermassive Black
  Hole Binaries on the Detection of Compact Binaries by LISA}. \emph{arXiv} {\bf 2023}.
\url{http://arxiv.org/abs/2309.14045}.

\bibitem{Evans:2023jia}
Evans, A.E.; Blecha, L.; Bhowmick, A.K.
\newblock {Building Semi-Analytic Black Hole Seeding Models Using IllustrisTNG
  Host Galaxies}. \emph{arXiv} {\bf 2023}.
\url{http://arxiv.org/abs/2309.11324}.

\bibitem{Gardiner:2023zzr}
Gardiner, E.C.; Kelley, L.Z.; Lemke, A.M.; Mitridate, A.
\newblock {Beyond the Background: Gravitational Wave Anisotropy and Continuous
  Waves from Supermassive Black Hole Binaries}. \emph{arXiv} {\bf 2023}.
\url{http://arxiv.org/abs/2309.07227}.

\bibitem{Serra:2023kkk}
Serra, F.
\newblock {Black Holes through the Lenses of Effective Field Theory}.
\newblock Ph.D. Thesis,  Scuola Normale Superiore,
  Pisa,  Italy, 2023.
\newblock {\url{https://doi.org/10.25429/serra-francesco_phd2023-09-13}}.

\bibitem{Cyr:2023pgw}
Cyr, B.; Kite, T.; Chluba, J.; Hill, J.C.; Jeong, D.; Acharya, S.K.; Bolliet,
  B.; Patil, S.P.
\newblock {Disentangling the primordial nature of stochastic gravitational wave
  backgrounds with CMB spectral distortions}. \emph{arXiv} {\bf 2023}.
\url{http://arxiv.org/abs/2309.02366}.

\bibitem{InternationalPulsarTimingArray:2023mzf}
Agazie, G.  et~al. [The International Pulsar Timing Array Collaboration].
\newblock {Comparing recent PTA results on the nanohertz stochastic
  gravitational wave background}. \emph{arXiv} {\bf 2023}.
\url{http://arxiv.org/abs/2309.00693}.

\bibitem{Flores:2023dgp}
Flores, M.M.; Kusenko, A.; Pearce, L.; Perez-Gonzalez, Y.F.; White, G.
\newblock {Testing high scale supersymmetry via second order gravitational
  waves}.
\newblock {\em Phys. Rev. D} {\bf 2023}, {\em 108},~123002.
\newblock {\url{https://doi.org/10.1103/PhysRevD.108.123002}}.

\bibitem{Kawasaki:2023rfx}
Kawasaki, M.; Murai, K.
\newblock {Enhancement of gravitational waves at Q-ball decay including
  non-linear density perturbations}.
\newblock {\em J. Cosmol. Astropart. Phys.} {\bf 2024}, {\em 01},~050.
\newblock {\url{https://doi.org/10.1088/1475-7516/2024/01/050}}.

\bibitem{Ellis:2023oxs}
Ellis, J.; Fairbairn, M.; Franciolini, G.; H\"utsi, G.; Iovino, A.; Lewicki,
  M.; Raidal, M.; Urrutia, J.; Vaskonen, V.; Veerm\"ae, H.
\newblock {What is the source of the PTA GW signal?}
\newblock {\em Phys. Rev. D} {\bf 2024}, {\em 109},~023522.
\newblock {\url{https://doi.org/10.1103/PhysRevD.109.023522}}.

\bibitem{Bhaumik:2023wmw}
Bhaumik, N.; Jain, R.K.; Lewicki, M.
\newblock {Ultralow mass primordial black holes in the early Universe can
  explain the pulsar timing array signal}.
\newblock {\em Phys. Rev. D} {\bf 2023}, {\em 108},~123532.
\newblock {\url{https://doi.org/10.1103/PhysRevD.108.123532}}.

\bibitem{Babichev:2023pbf}
Babichev, E.; Gorbunov, D.; Ramazanov, S.; Samanta, R.; Vikman, A.
\newblock {NANOGrav spectral index \ensuremath{\gamma}=3 from melting domain
  walls}.
\newblock {\em Phys. Rev. D} {\bf 2023}, {\em 108},~123529.
\newblock {\url{https://doi.org/10.1103/PhysRevD.108.123529}}.

\bibitem{Buchmuller:2023aus}
Buchmuller, W.; Domcke, V.; Schmitz, K.
\newblock {Metastable cosmic strings}.
\newblock {\em J. Cosmol. Astropart. Phys.} {\bf 2023}, {\em 11},~020.
\newblock {\url{https://doi.org/10.1088/1475-7516/2023/11/020}}.

\bibitem{Gouttenoire:2023bqy}
Gouttenoire, Y.
\newblock {First-Order Phase Transition Interpretation of Pulsar Timing Array
  Signal Is Consistent with Solar-Mass Black Holes}.
\newblock {\em Phys. Rev. Lett.} {\bf 2023}, {\em 131},~171404.
\newblock {\url{https://doi.org/10.1103/PhysRevLett.131.171404}}.

\bibitem{Wu:2023hsa}
Wu, Y.M.; Chen, Z.C.; Huang, Q.G.
\newblock {Cosmological Interpretation for the Stochastic Signal in Pulsar
  Timing Arrays}. \emph{arXiv} {\bf 2023}.
\url{http://arxiv.org/abs/2307.03141}.

\bibitem{Zhang:2023lzt}
Zhang, C.; Dai, N.; Gao, Q.; Gong, Y.; Jiang, T.; Lu, X.
\newblock {Detecting new fundamental fields with pulsar timing arrays}.
\newblock {\em Phys. Rev. D} {\bf 2023}, {\em 108},~104069.
\newblock {\url{https://doi.org/10.1103/PhysRevD.108.104069}}.

\bibitem{Gouttenoire:2023nzr}
Gouttenoire, Y.; Trifinopoulos, S.; Valogiannis, G.; Vanvlasselaer, M.
\newblock {Scrutinizing the Primordial Black Holes Interpretation of PTA
  Gravitational Waves and JWST Early Galaxies}. \emph{arXiv} {\bf 2023}.
\url{http://arxiv.org/abs/2307.01457}.

\bibitem{Bi:2023tib}
Bi, Y.C.; Wu, Y.M.; Chen, Z.C.; Huang, Q.G.
\newblock {Implications for the supermassive black hole binaries from the
  NANOGrav 15-year data set}.
\newblock {\em Sci. China Phys. Mech. Astron.} {\bf 2023}, {\em 66},~120402.
\newblock {\url{https://doi.org/10.1007/s11433-023-2252-4}}.

\bibitem{Broadhurst:2023tus}
Broadhurst, T.; Chen, C.; Liu, T.; Zheng, K.F.
\newblock {Binary Supermassive Black Holes Orbiting Dark Matter Solitons: From
  the Dual AGN in UGC4211 to NanoHertz Gravitational Waves}. \emph{arXiv} {\bf 2023}.
\url{http://arxiv.org/abs/2306.17821}.

\bibitem{Huang:2023chx}
Huang, H.L.; Cai, Y.; Jiang, J.Q.; Zhang, J.; Piao, Y.S.
\newblock {Supermassive primordial black holes in multiverse: for nano-Hertz
  gravitational wave and high-redshift JWST galaxies}. \emph{arXiv} {\bf 2023}.
\url{http://arxiv.org/abs/2306.17577}.

\bibitem{Ellis:2023tsl}
Ellis, J.; Lewicki, M.; Lin, C.; Vaskonen, V.
\newblock {Cosmic superstrings revisited in light of NANOGrav 15-year data}.
\newblock {\em Phys. Rev. D} {\bf 2023}, {\em 108},~103511.
\newblock {\url{https://doi.org/10.1103/PhysRevD.108.103511}}.

\bibitem{Ellis:2023dgf}
Ellis, J.; Fairbairn, M.; H\"utsi, G.; Raidal, J.; Urrutia, J.; Vaskonen, V.;
  Veerm\"ae, H.
\newblock {Gravitational waves from supermassive black hole binaries in light
  of the NANOGrav 15-year data}.
\newblock {\em Phys. Rev. D} {\bf 2024}, {\em 109},~L021302.
\newblock {\url{https://doi.org/10.1103/PhysRevD.109.L021302}}.

\bibitem{Addazi:2023jvg}
Addazi, A.; Cai, Y.F.; Marciano, A.; Visinelli, L.
\newblock {Have pulsar timing array methods detected a cosmological phase
  transition?}
\newblock {\em Phys. Rev. D} {\bf 2024}, {\em 109},~015028.
\newblock {\url{https://doi.org/10.1103/PhysRevD.109.015028}}.

\bibitem{Furusawa:2023fwl}
Furusawa, K.; Tashiro, H.; Yokoyama, S.; Ichiki, K.
\newblock {Probing the Mass Relation between Supermassive Black Holes and Dark
  Matter Halos at High Redshifts by Gravitational Wave Experiments}.
\newblock {\em Astrophys. J.} {\bf 2023}, {\em 959},~117.
\newblock {\url{https://doi.org/10.3847/1538-4357/ad088f}}.

\bibitem{Chen:2024fir}
Chen, Z.C.; Li, J.; Liu, L.; Yi, Z.
\newblock {Probing the speed of scalar-induced gravitational waves with pulsar
  timing arrays}. \emph{arXiv} {\bf 2024}.
\newblock  \href{http://arxiv.org/abs/2401.09818}{{\normalfont
  [arXiv:gr-qc/2401.09818]}}.

\bibitem{Li:2024psa}
Li, H.J.; Zhou, Y.F.
\newblock {Gravitational waves from axion domain walls in double level
  crossings}. \emph{arXiv} {\bf 2024}.
\newblock  \href{http://arxiv.org/abs/2401.09138}{{\normalfont
  [arXiv:hep-ph/2401.09138]}}.

\bibitem{Liu:2023tmv}
Liu, J.
\newblock {Distinguishing the nanohertz gravitational-wave sources by the
  observations of compact dark matter subhalos}.
\newblock {\em Phys. Rev. D} {\bf 2023}, {\em 108},~123544.
\newblock {\url{https://doi.org/10.1103/PhysRevD.108.123544}}.

\bibitem{Chen:2023bms}
Chen, Z.C.; Li, S.L.; Wu, P.; Yu, H.
\newblock {NANOGrav hints for first-order confinement-deconfinement phase
  transition in different QCD-matter scenarios}. \emph{arXiv} {\bf 2023}.
\url{http://arxiv.org/abs/2312.01824}.

\bibitem{Kitajima:2023kzu}
Kitajima, N.; Lee, J.; Takahashi, F.; Yin, W.
\newblock {Stability of domain walls with inflationary fluctuations under
  potential bias, and gravitational wave signatures}. \emph{arXiv} {\bf 2023}.
\url{http://arxiv.org/abs/2311.14590}.

\bibitem{LISACosmologyWorkingGroup:2023njw}
Bagui, E.  et~al. [LISA Cosmology Working Group].
\newblock {Primordial black holes and their gravitational-wave signatures}. \emph{arXiv} {\bf
  2023}.
\url{http://arxiv.org/abs/2310.19857}.

\bibitem{Liu:2023hpw}
Liu, L.; Wu, Y.; Chen, Z.C.
\newblock {Simultaneously probing the sound speed and equation of state of the
  early Universe with pulsar timing arrays}. \emph{arXiv} {\bf 2023}.
\url{http://arxiv.org/abs/2310.16500}.

\bibitem{Chung:2023xcv}
Chung, D.J.H.; Tadepalli, S.C.
\newblock {Power spectrum in the chaotic regime of axionic blue isocurvature
  perturbations}.
\newblock {\em Phys. Rev. D} {\bf 2024}, {\em 109},~023539.
\newblock {\url{https://doi.org/10.1103/PhysRevD.109.023539}}.

\bibitem{King:2023ayw}
King, S.F.; Roshan, R.; Wang, X.; White, G.; Yamazaki, M.
\newblock {Quantum gravity effects on dark matter and gravitational waves}.
\newblock {\em Phys. Rev. D} {\bf 2024}, {\em 109},~024057.
\newblock {\url{https://doi.org/10.1103/PhysRevD.109.024057}}.

\bibitem{Zhu:2023lbf}
Zhu, M.; Ye, G.; Cai, Y.
\newblock {Pulsar timing array observations as possible hints for nonsingular
  cosmology}.
\newblock {\em Eur. Phys. J. C} {\bf 2023}, {\em 83},~816.
\newblock {\url{https://doi.org/10.1140/epjc/s10052-023-11963-4}}.

\bibitem{Liu:2023pau}
Liu, L.; Chen, Z.C.; Huang, Q.G.
\newblock {Probing the equation of state of the early Universe with pulsar
  timing arrays}.
\newblock {\em J. Cosmol. Astropart. Phys.} {\bf 2023}, {\em 11},~071.
\newblock {\url{https://doi.org/10.1088/1475-7516/2023/11/071}}.

\bibitem{Ahmadvand:2023lpp}
Ahmadvand, M.; Bian, L.; Shakeri, S.
\newblock {Heavy QCD axion model in light of pulsar timing arrays}.
\newblock {\em Phys. Rev. D} {\bf 2023}, {\em 108},~115020.
\newblock {\url{https://doi.org/10.1103/PhysRevD.108.115020}}.

\bibitem{Zhang:2023nrs}
Zhang, Z.; Cai, C.; Su, Y.H.; Wang, S.; Yu, Z.H.; Zhang, H.H.
\newblock {Nano-Hertz gravitational waves from collapsing domain walls
  associated with freeze-in dark matter in light of pulsar timing array
  observations}.
\newblock {\em Phys. Rev. D} {\bf 2023}, {\em 108},~095037.
\newblock {\url{https://doi.org/10.1103/PhysRevD.108.095037}}.

\bibitem{Cannizzaro:2023mgc}
Cannizzaro, E.; Franciolini, G.; Pani, P.
\newblock {Novel tests of gravity using nano-Hertz stochastic
  gravitational-wave background signals}. \emph{arXiv} {\bf 2023}.
\url{http://arxiv.org/abs/2307.11665}.

\bibitem{Jin:2023wri}
Jin, J.H.; Chen, Z.C.; Yi, Z.; You, Z.Q.; Liu, L.; Wu, Y.
\newblock {Confronting sound speed resonance with pulsar timing arrays}.
\newblock {\em J. Cosmol. Astropart. Phys.} {\bf 2023}, {\em 09},~016.
\newblock {\url{https://doi.org/10.1088/1475-7516/2023/09/016}}.

\bibitem{Servant:2023mwt}
Servant, G.; Simakachorn, P.
\newblock {Constraining postinflationary axions with pulsar timing arrays}.
\newblock {\em Phys. Rev. D} {\bf 2023}, {\em 108},~123516.
\newblock {\url{https://doi.org/10.1103/PhysRevD.108.123516}}.

\bibitem{Li:2023bxy}
Li, S.P.; Xie, K.P.
\newblock {Collider test of nano-Hertz gravitational waves from pulsar timing
  arrays}.
\newblock {\em Phys. Rev. D} {\bf 2023}, {\em 108},~055018.
\newblock {\url{https://doi.org/10.1103/PhysRevD.108.055018}}.

\bibitem{Gouttenoire:2023ftk}
Gouttenoire, Y.; Vitagliano, E.
\newblock {Domain wall interpretation of the PTA signal confronting black hole
  overproduction}. \emph{arXiv} {\bf 2023}.
\url{http://arxiv.org/abs/2306.17841}.

\bibitem{Blasi:2023sej}
Blasi, S.; Mariotti, A.; Rase, A.; Sevrin, A.
\newblock {Axionic domain walls at Pulsar Timing Arrays: QCD bias and particle
  friction}.
\newblock {\em JHEP} {\bf 2023}, {\em 11},~169.
\newblock {\url{https://doi.org/10.1007/JHEP11(2023)169}}.

\bibitem{Vagnozzi:2023lwo}
Vagnozzi, S.
\newblock {Inflationary interpretation of the stochastic gravitational wave
  background signal detected by pulsar timing array experiments}.
\newblock {\em  J. High Energy Astrophys.} {\bf 2023}, {\em 39},~81--98.
\newblock {\url{https://doi.org/10.1016/j.jheap.2023.07.001}}.

\bibitem{Franciolini:2023pbf}
Franciolini, G.; Iovino, Junior., A.; Vaskonen, V.; Veermae, H.
\newblock {Recent Gravitational Wave Observation by Pulsar Timing Arrays and
  Primordial Black Holes: The Importance of Non--Gaussianities}.
\newblock {\em Phys. Rev. Lett.} {\bf 2023}, {\em 131},~201401.
\newblock {\url{https://doi.org/10.1103/PhysRevLett.131.201401}}.

\bibitem{Athron:2023mer}
Athron, P.; Fowlie, A.; Lu, C.T.; Morris, L.; Wu, L.; Wu, Y.; Xu, Z.
\newblock {Can supercooled phase transitions explain the gravitational wave
  background observed by pulsar timing arrays?} \emph{arXiv} {\bf 2023}.
\url{http://arxiv.org/abs/2306.17239}.

\bibitem{Zeng:2023jut}
Zeng, Z.M.; Liu, J.; Guo, Z.K.
\newblock {Enhanced curvature perturbations from spherical domain walls
  nucleated during inflation}.
\newblock {\em Phys. Rev. D} {\bf 2023}, {\em 108},~063005.
\newblock {\url{https://doi.org/10.1103/PhysRevD.108.063005}}.

\bibitem{Gonzalez:2022mcx}
Gonzalez, D.; Kitajima, N.; Takahashi, F.; Yin, W.
\newblock {Stability of domain wall network with initial inflationary
  fluctuations and its implications for cosmic birefringence}.
\newblock {\em Phys. Lett. B} {\bf 2023}, {\em 843},~137990.
\newblock {\url{https://doi.org/10.1016/j.physletb.2023.137990}}.

\bibitem{Blasi:2022ayo}
Blasi, S.; Mariotti, A.; Rase, A.; Sevrin, A.; Turbang, K.
\newblock {Friction on ALP domain walls and gravitational waves}.
\newblock {\em J. Cosmol. Astropart. Phys.} {\bf 2023}, {\em 04},~008.
\newblock {\url{https://doi.org/10.1088/1475-7516/2023/04/008}}.

\bibitem{Chattopadhyay:2022fwa}
Chattopadhyay, P.; Chaudhuri, A.; Khlopov, M.Y.
\newblock {Dark Matter from Evaporating PBH dominated in the Early Universe}. \emph{arXiv}
  {\bf 2022}.
\newblock  \href{http://arxiv.org/abs/2209.11288}{{\normalfont
  [arXiv:hep-ph/2209.11288]}}.

\bibitem{Wu:2022tpe}
Wu, Y.; Xie, K.P.; Zhou, Y.L.
\newblock {Classification of Abelian domain walls}.
\newblock {\em Phys. Rev. D} {\bf 2022}, {\em 106},~075019.
\newblock {\url{https://doi.org/10.1103/PhysRevD.106.075019}}.

\bibitem{Ferreira:2022zzo}
Ferreira, R.Z.; Notari, A.; Pujolas, O.; Rompineve, F.
\newblock {Gravitational waves from domain walls in Pulsar Timing Array
  datasets}.
\newblock {\em J. Cosmol. Astropart. Phys.} {\bf 2023}, {\em 02},~001.
\newblock {\url{https://doi.org/10.1088/1475-7516/2023/02/001}}.

\bibitem{Wu:2022stu}
Wu, Y.; Xie, K.P.; Zhou, Y.L.
\newblock {Collapsing domain walls beyond Z2}.
\newblock {\em Phys. Rev. D} {\bf 2022}, {\em 105},~095013.
\newblock {\url{https://doi.org/10.1103/PhysRevD.105.095013}}.

\bibitem{Ashoorioon:2022raz}
Ashoorioon, A.; Rezazadeh, K.; Rostami, A.
\newblock {NANOGrav signal from the end of inflation and the LIGO mass and
  heavier primordial black holes}.
\newblock {\em Phys. Lett. B} {\bf 2022}, {\em 835},~137542.
\newblock {\url{https://doi.org/10.1016/j.physletb.2022.137542}}.

\bibitem{Sun:2021yra}
Sun, S.; Yang, X.Y.; Zhang, Y.L.
\newblock {Pulsar timing residual induced by wideband ultralight dark matter
  with spin 0,1,2}.
\newblock {\em Phys. Rev. D} {\bf 2022}, {\em 106},~066006.
\newblock {\url{https://doi.org/10.1103/PhysRevD.106.066006}}.

\bibitem{Babichev:2021uvl}
Babichev, E.; Gorbunov, D.; Ramazanov, S.; Vikman, A.
\newblock {Gravitational shine of dark domain walls}.
\newblock {\em J. Cosmol. Astropart. Phys.} {\bf 2022}, {\em 04},~028.
\newblock {\url{https://doi.org/10.1088/1475-7516/2022/04/028}}.

\bibitem{Benetti:2021uea}
Benetti, M.; Graef, L.L.; Vagnozzi, S.
\newblock {Primordial gravitational waves from NANOGrav: A broken power-law
  approach}.
\newblock {\em Phys. Rev. D} {\bf 2022}, {\em 105},~043520.
\newblock {\url{https://doi.org/10.1103/PhysRevD.105.043520}}.

\bibitem{Kirillov:2021qjz}
Kirillov, A.A.; Rubin, S.G.
\newblock {On Mass Spectra of Primordial Black Holes}.
\newblock {\em Front. Astron. Space Sci.} {\bf 2021}, {\em 8},~777661.
\newblock {\url{https://doi.org/10.3389/fspas.2021.777661}}.

\bibitem{Brandenburg:2021tmp}
Brandenburg, A.; Clarke, E.; He, Y.; Kahniashvili, T.
\newblock {Can we observe the QCD phase transition-generated gravitational
  waves through pulsar timing arrays?}
\newblock {\em Phys. Rev. D} {\bf 2021}, {\em 104},~043513.
\newblock {\url{https://doi.org/10.1103/PhysRevD.104.043513}}.

\bibitem{Sakharov:2021dim}
Sakharov, A.S.; Eroshenko, Y.N.; Rubin, S.G.
\newblock {Looking at the NANOGrav signal through the anthropic window of
  axionlike particles}.
\newblock {\em Phys. Rev. D} {\bf 2021}, {\em 104},~043005.
\newblock {\url{https://doi.org/10.1103/PhysRevD.104.043005}}.

\bibitem{NANOGrav:2020bcs}
Arzoumanian, Z.  et~al. [The NANOGrav Collaboration].
\newblock {The NANOGrav 12.5 yr Data Set: Search for an Isotropic Stochastic
  Gravitational-wave Background}.
\newblock {\em Astrophys. J. Lett.} {\bf 2020}, {\em 905},~L34.
\newblock {\url{https://doi.org/10.3847/2041-8213/abd401}}.

\bibitem{2007PhLB..651..224N}
{Nojiri}, S.; {Odintsov}, S.D.; {Tretyakov}, P.V.
\newblock {Dark energy from modified F(R)-scalar-Gauss Bonnet gravity}.
\newblock {\em Phys. Lett. B} {\bf 2007}, {\em 651},~224--231.
\newblock {\url{https://doi.org/10.1016/j.physletb.2007.06.029}}.

\bibitem{Nojiri_2017}
Nojiri, S.; Odintsov, S.D.; Oikonomou, V.K.
\newblock {Modified Gravity Theories on a Nutshell: Inflation, Bounce and
  Late-time Evolution}.
\newblock {\em Phys. Rept.} {\bf 2017}, {\em 692},~1--104.
\newblock {\url{https://doi.org/10.1016/j.physrep.2017.06.001}}.

\bibitem{Rubin:2015pqa}
Rubin, S.G.
\newblock {Scalar field localization on deformed extra space}.
\newblock {\em Eur. Phys. J.} {\bf 2015}, {\em C75},~333.
\newblock {\url{https://doi.org/10.1140/epjc/s10052-015-3553-z}}.

\bibitem{2017JCAP...10..001B}
{Bronnikov}, K.A.; {Budaev}, R.I.; {Grobov}, A.V.; {Dmitriev}, A.E.; {Rubin},
  S.G.
\newblock {Inhomogeneous compact extra dimensions}.
\newblock {\em J. Cosmol. Astropart. Phys.} {\bf 2017}, {\em 10},~001.
\newblock {\url{https://doi.org/10.1088/1475-7516/2017/10/001}}.

\bibitem{Rubin:2016ude}
Rubin, S.G.
\newblock {Inhomogeneous extra space as a tool for the top-down approach}.
\newblock {\em Adv. High Energy Phys.} {\bf 2018}, {\em 2018},~2767410.
\newblock {\url{https://doi.org/10.1155/2018/2767410}}.

\bibitem{Bronnikov:2023lej}
Bronnikov, K.A.; Popov, A.A.; Rubin, S.G.
\newblock {Multi-scale hierarchy from multidimensional gravity}.
\newblock {\em Phys. Dark Univ.} {\bf 2023}, {\em 42},~101378.
\newblock {\url{https://doi.org/10.1016/j.dark.2023.101378}}.

\bibitem{Nojiri:2006ri}
Nojiri, S.; Odintsov, S.D.
\newblock {Introduction to modified gravity and gravitational alternative for
  dark energy}.
\newblock {\em eConf} {\bf 2006}, {\em C0602061},~06.
\newblock {\url{https://doi.org/10.1142/S0219887807001928}}.

\bibitem{Sotiriou:2008rp}
Sotiriou, T.P.; Faraoni, V.
\newblock {f(R) Theories Of Gravity}.
\newblock {\em Rev. Mod. Phys.} {\bf 2010}, {\em 82},~451--497,
  \href{http://arxiv.org/abs/0805.1726}{{\normalfont [arXiv:gr-qc/0805.1726]}}.
\newblock {\url{https://doi.org/10.1103/RevModPhys.82.451}}.

\bibitem{Arbuzova:2021etq}
Arbuzova, E.; Dolgov, A.; Singh, R.
\newblock {$R^2$-Cosmology and New Windows for Superheavy Dark Matter}.
\newblock {\em Symmetry} {\bf 2021}, {\em 13},~877.
\newblock {\url{https://doi.org/10.3390/sym13050877}}.

\bibitem{DeFelice:2010aj}
De~Felice, A.; Tsujikawa, S.
\newblock {f(R) theories}.
\newblock {\em Living Rev. Rel.} {\bf 2010}, {\em 13},~3.
\newblock {\url{https://doi.org/10.12942/lrr-2010-3}}.

\bibitem{2014JCAP...01..008B}
{Bamba}, K.; {Makarenko}, A.N.; {Myagky}, A.N.; {Nojiri}, S.; {Odintsov}, S.D.
\newblock {Bounce cosmology from F(R) gravity and F(R) bigravity}.
\newblock {\em J. Cosmol. Astropart. Phys.} {\bf 2014}, {\em 1},~8.
\newblock {\url{https://doi.org/10.1088/1475-7516/2014/01/008}}.

\bibitem{Sokolowski:2007rd}
Sokolowski, L.M.
\newblock {Metric gravity theories and cosmology:II. Stability of a ground
  state in f(R) theories}.
\newblock {\em Class. Quant. Grav.} {\bf 2007}, {\em 24},~3713--3734.
\newblock {\url{https://doi.org/10.1088/0264-9381/24/14/011}}.

\bibitem{2000PhRvD..62d4014O}
{Olasagasti}, I.; {Vilenkin}, A.
\newblock {Gravity of higher-dimensional global defects}.
\newblock {\em Phys. Rev.~{{D}}} {\bf 2000}, {\em 62},~044014.
\newblock {\url{https://doi.org/10.1103/PhysRevD.62.044014}}.

\bibitem{2003PhRvD..68b5013C}
{Cho}, I.; {Vilenkin}, A.
\newblock {Gravity of superheavy higher-dimensional global defects}.
\newblock {\em Phys. Rev.~{{D}}} {\bf 2003}, {\em 68},~025013.
\newblock {\url{https://doi.org/10.1103/PhysRevD.68.025013}}.

\bibitem{2005PhRvD..71h4002S}
{Shimono}, S.; {Chiba}, T.
\newblock {Numerical solutions of inflating higher dimensional global defects}.
\newblock {\em Phys. Rev.~{{D}}} {\bf 2005}, {\em 71},~084002.
\newblock {\url{https://doi.org/10.1103/PhysRevD.71.084002}}.

\bibitem{2005PhRvD..71j4018R}
{Ringeval}, C.; {Peter}, P.; {Uzan}, J.P.
\newblock {Stability of six-dimensional hyperstring braneworlds}.
\newblock {\em Phys. Rev.~{{D}}} {\bf 2005}, {\em 71},~104018.
\newblock {\url{https://doi.org/10.1103/PhysRevD.71.104018}}.

\bibitem{2000PhRvL..84.2564G}
{Gregory}, R.
\newblock {Nonsingular Global String Compactifications}.
\newblock {\em Phys. Rev. Lett.} {\bf 2000}, {\em 84},~2564--2567.
\newblock {\url{https://doi.org/10.1103/PhysRevLett.84.2564}}.

\bibitem{2000PhRvL..85..240G}
{Gherghetta}, T.; {Shaposhnikov}, M.
\newblock {Localizing Gravity on a Stringlike Defect in Six Dimensions}.
\newblock {\em Phys. Rev. Lett.} {\bf 2000}, {\em 85},~240--243.
\newblock {\url{https://doi.org/10.1103/PhysRevLett.85.240}}.

\bibitem{Bronnikov:2007kw}
Bronnikov, K.; Meierovich, B.
\newblock {Global strings in extra dimensions: A full map of solutions, matter
  trapping and the hierarchy problem}.
\newblock {\em J.\ Exp.\ Theor.\ Phys.} {\bf 2008}, {\em 106},~247--264.
\newblock {\url{https://doi.org/10.1007/s11447-008-2005-0}}.

\bibitem{Planck:Hinfl}
Akrami, Y.  et~al. [Planck Collaboration].
\newblock {Planck 2018 results. X. Constraints on inflation}.
\newblock {\em Astron. Astrophys.} {\bf 2020}, {\em 641},~A10.
\newblock {\url{https://doi.org/10.1051/0004-6361/201833887}}.

\bibitem{clusterGW1}
Depta, P.F.; Schmidt-Hoberg, K.; Schwaller, P.; Tasillo, C.
\newblock {Do pulsar timing arrays observe merging primordial black holes?}  \emph{arXiv}
  {\bf 2023}.
\url{http://arxiv.org/abs/2306.17836}.

\bibitem{2021arXiv210908373R}
{Rubin}, S.G.; {Fabris}, J.C.
\newblock {Distortion of extra dimensions in the inflationary Multiverse}.
\newblock {\em arXiv} {\bf 2021}.
  \url{http://arxiv.org/abs/2109.08373}.

\end{thebibliography}
\end{document}